\documentclass[%
 reprint,
 amsmath,amssymb,
 aps,
 prb,
nobalancelastpage
]{revtex4-2}

\usepackage{graphicx}
\usepackage{dcolumn}
\usepackage{bm}
\usepackage{xcolor}
\usepackage{lineno}
\usepackage{hyperref}

\hypersetup{
    colorlinks=true,
    linkcolor=blue,
    filecolor=magenta,      
    urlcolor=cyan,
}

\newcommand{\cc}{\hat{c}}
\newcommand{\cdag}{\hat{c}^\dagger}
\newcommand{\mm}{{\bf m}}
\newcommand{\nn}{{\bf n}}
\newcommand{\RR}{{\bf R}}
\newcommand{\TT}{{\bf T}}

\newcommand{\rr}{{\bf r}}
\newcommand{\kk}{{\bf k}}
\newcommand{\qq}{{\bf q}}
\newcommand{\kq}{{\bf k+q}}
\newcommand{\GG}{{\bf G}}
\newcommand{\btau}{{\bm \tau}}
\newcommand{\zij}{z_{ij}}
\newcommand{\rij}{r_{ij}}
\newcommand{\tij}{t_{ij}}
\newcommand{\g}{\hat{g}}

\newcommand*\xbar[1]{%
   \hbox{%
     \vbox{%
       \hrule height 0.5pt 
       \kern0.5ex
       \hbox{%
         \kern-0.1em
         \ensuremath{#1}%
         \kern-0.1em
       }%
     }%
   }%
} 
\newcommand*\ybar[1]{%
   \hbox{%
     \scriptsize
     \vbox{%
       \hrule height 0.5pt 
       \kern0.3ex
       \hbox{%
         \kern-0.1em
         \ensuremath{#1}%
         \kern-0.1em
       }%
     }%
   }%
} 

\begin{document}

\title{Construction of low-energy symmetric Hamiltonians and Hubbard parameters for twisted multilayer systems using \textit{ab-initio} input}

\author{Arkadiy Davydov}
\author{Kenny Choo}
\author{Mark H. Fischer}
\author{Titus Neupert}
\affiliation{Department of Physics, University of Zurich, Winterthurerstrasse 190, 8057 Zurich, Switzerland}

\date{\today}

\begin{abstract}
A computationally efficient workflow for obtaining the low-energy symmetric tight-binding Hamiltonians for twisted multilayer systems is presented in this work. We apply this scheme to twisted bilayer graphene at the first magic angle. As initial step, the full-energy tight-binding Hamiltonian is generated by the Slater-Koster model with parameters fitted to \textit{ab-initio} data at larger angles. Then, the low-energy symmetric four-band and twelve-band Hamiltonians are constructed using the maximum-localization procedure subjected to crystal and time-reversal-symmetry constraints. Finally, we compute extended Hubbard parameters for both models within the constrained random phase approximation (cRPA) for screening, which again respect the symmetries. The relevant data and results of this work are freely available via an online repository. Our workflow, exemplified in this work on twisted bilayer graphene, is straightforwardly transferable to other twisted multi-layer materials.
\end{abstract}

\maketitle

\section{Introduction}

The discovery of correlated insulating states and superconductivity in twisted bilayer graphene (TBG)~\cite{Cao2018_TBG_correlated, Cao2018_TBG_superconductivity,Kim2017} has rapidly opened the field of twistronics with stacked van der Waals materials. In TBG at the so-called first magic twist angle $1.05^\circ$,  nearly flat bands around the Fermi level emerge, which is the key ingredient for strongly-correlated states to emerge in an otherwise weakly-correlated material~\cite{Cao2018_TBG_correlated,Xie2019,Yankowitz2019,Kerelsky2019,Lu2019,Lisi2021}. The observation of a correlated insulator in close proximity to superconductivity further leads one to believe that the superconductivity is of an unconventional nature. 

For a theoretical investigation of the phenomena observed in TBG, two ingredients are essential: (1) a faithful description of the low-energy band structure and (2) the form and approximate strength of the relevant interactions. However, performing \textit{ab-initio} calculations---a standard procedure for finding effective models, such as tight-binding (TB) models---is challenging as the unit cell at the magic angle of $1.05^{\circ}$ contains about twelve thousand atoms, which is prohibitively large for such first-principles calculations. While \textit{ab-initio} calculations to obtain the full bandstructure were carried out in Ref.~\cite{PhysRevB.99.195419}, the required computational cost places strong constraints on the pseudopotentials and the computational flow. Therefore, simplified schemes for the derivation of effective models in twisted heterostructures are highly desirable. 

The simplest approach to describe the (non-interacting) electronic states of TBG is given by the so-called continuum theory or $\bf{k}\cdot\bf{p}$ approximation~\cite{LopesDosSantos2007,Koshino2018,Bistritzer2011,Bernevig2020tbg1}. Starting from the linearly dispersing Hamiltonian of individual graphene sheets and adding interlayer couplings, the resulting models give a reasonable approximation to the electronic energy dispersion of TBG at arbitrary angles. In addition, the models respect the $D_6$ symmetry group which protects the Dirac points~\cite{Zou2018}. Even though the $\bf{k}\cdot\bf{p}$ method is straight forward to implement, the resulting models ignore quantitative microscopic details, which may alter the qualitative nature of electronic states, including their symmetry or topology.

Another approach is to directly start from a microscopic TB model as an approximate but compact representation of the full, experimentally realistic Hamiltonian in a basis of localized Wannier orbitals. For graphene, the orbital character of states near the Fermi level is dominated by $p_z$ orbitals, which remains true in TBG. For fixed orbital shape and orientation, the standard method for obtaining the hopping parameters of the TB Hamiltonian is to assume the Slater-Koster (SK) analytic form with standard values for the SK parameters~\cite{Koshino2018,TramblyDeLaissardiere2012,DeTramblyLaissardiere2010,Goodwin2019_cRPA,Goodwin2019,Haddadi2020}. While this approach allows one to describe systems with unit-cell sizes relevant for small twist angles, the fixed choice of SK parameters restricts the overall accuracy of the scheme.

In this work, we consider a compromise between accuracy and computational cost. Instead of computing the \textit{ab-initio} TB model at the magic angle, we first optimize the SK parameters according to the \textit{ab-initio} TB parameters calculated at larger twist angles. The resulting SK parameters are then used for a TB description for TBG at smaller twist angles. In other words, the SK analytic form is used to extrapolate \textit{ab-initio} information from larger angles to smaller angles. This procedure allows for a significant reduction in computational cost due to the smaller unit cells. We verify the consistency of our approach by comparing the SK parameters obtained using different twist angles. Note that this scheme can also be applied to other twisted multi-layer materials. 

The full TB description is not directly amenable towards a many-body analysis due to the huge number of degrees of freedom stemming from the large unit cell close to the magic angle.
Therefore, the next step is to construct a minimal, non-interacting low-energy Hamiltonian, which describes the few relevant bands near the Fermi level. 
We build the low-energy TB Hamiltonians from the SK bandstructures, using the projection method adopted in the Wannier90 software~\cite{Pizzi2020}. Since the flat bands of TBG are isolated from others, and a good trial orbital set to initiate the procedure is known~\cite{Kang2019}, a four-band TB model can, thus, be easily obtained. 

Even at the magic angle the gap between the flat bands to other bands is small  (experimentally, $\sim 35$ meV~\cite{Zou2018}) compared to, for example, the phonon bandwidth ($\sim 200$ meV~\cite{Choi2018}). Thus, to study electron-phonon interactions it is desirable to include additional bands close to the Fermi level.
Without a gap separating these additional bands from the rest of the spectrum, the Wannierization requires an appropriate `disentanglement' step to separate the Hilbert space of the desired energy range from that of the rest of the band structure. This construction of a low-energy Hamiltonian is non-trivial, though recently, a solution was proposed requiring a multi-step projection procedure to obtain a good Wannier basis~\cite{PhysRevResearch.1.033072}. In this work, we report an alternative and simpler procedure to obtain a twelve-band low-energy Hamiltonian, using the standard routines available in the Wannier90 package~\cite{Pizzi2020}, and a simple modification of the code to include time-reversal (TR) symmetry. We further provide the Wyckoff positions and symmetry representations of Wannier orbitals required for stable convergence of the crystal-symmetry related Wannier90 subroutines and the Wannierization procedure itself.

Note that two possible configurations of commensurate unit cells for TBG exist: The unit cells possess either $D_3$ or $D_6$ symmetry, obtained by rotating $AA$-stacked graphene sheets around carbon centers or hexagon centers, respectively. In the latter case, a Wannier obstruction was reported, which would hinder the construction of a four-band tight-binding Hamiltonian with well-defined symmetry transformation of the basis Wannier functions due to the fragile topological nature of the flat bands~\cite{Zou2018,Po2018_origin_of_mott,Po2018_fragile_topo,Po2019}. In the $D_3$-symmetric case, which we focus on in this article, no such obstruction exists even in the single-valley theories, and a well-localized, symmetric four-band TB model can be obtained~\footnote{Note that a successful Wannierization of narrow bands of TBG within the continuum model was reported recently~\cite{Cao2021}, where both TBG valleys were considered. Therefore, the fragile topological nature of the low-energy Wannier bands may be an artifact of the single-valley approximation, and, therefore, the $D_6$-symmetric models can be obtained in the future by the proposed workflow as well.}.

To provide not only the single-particle Hamiltonian but also electron-electron interactions, we further require the form of the extended Hubbard parameters \cite{Koshino2018,Goodwin2019_cRPA,Goodwin2019,bernevig2020tbg3}. To take into account electronic screening, we make the constrained random phase approximation (cRPA), in which the considered low-energy manifold is excluded from the screening process.  We compute these parameters for both a four-band and twelve-band low-energy model, and discuss the symmetry constraints they obey.

The remainder of this article is organized as follows: In Sec.~\ref{sec:tbmodels}, the full workflow to construct a low-energy effective tight-binding model starting from the extraction of the SK parameters to the projection to four and twelve bands is described.
Section~\ref{sec:interactions} describes the cRPA calculation and Sec.~\ref{sec:results} shows the results, while Sec.~\ref{sec:conclusion} concludes our work.

\section{Construction of Tight-binding models}\label{sec:tbmodels}
\begin{figure}
\centering
\includegraphics[width=0.47\textwidth]{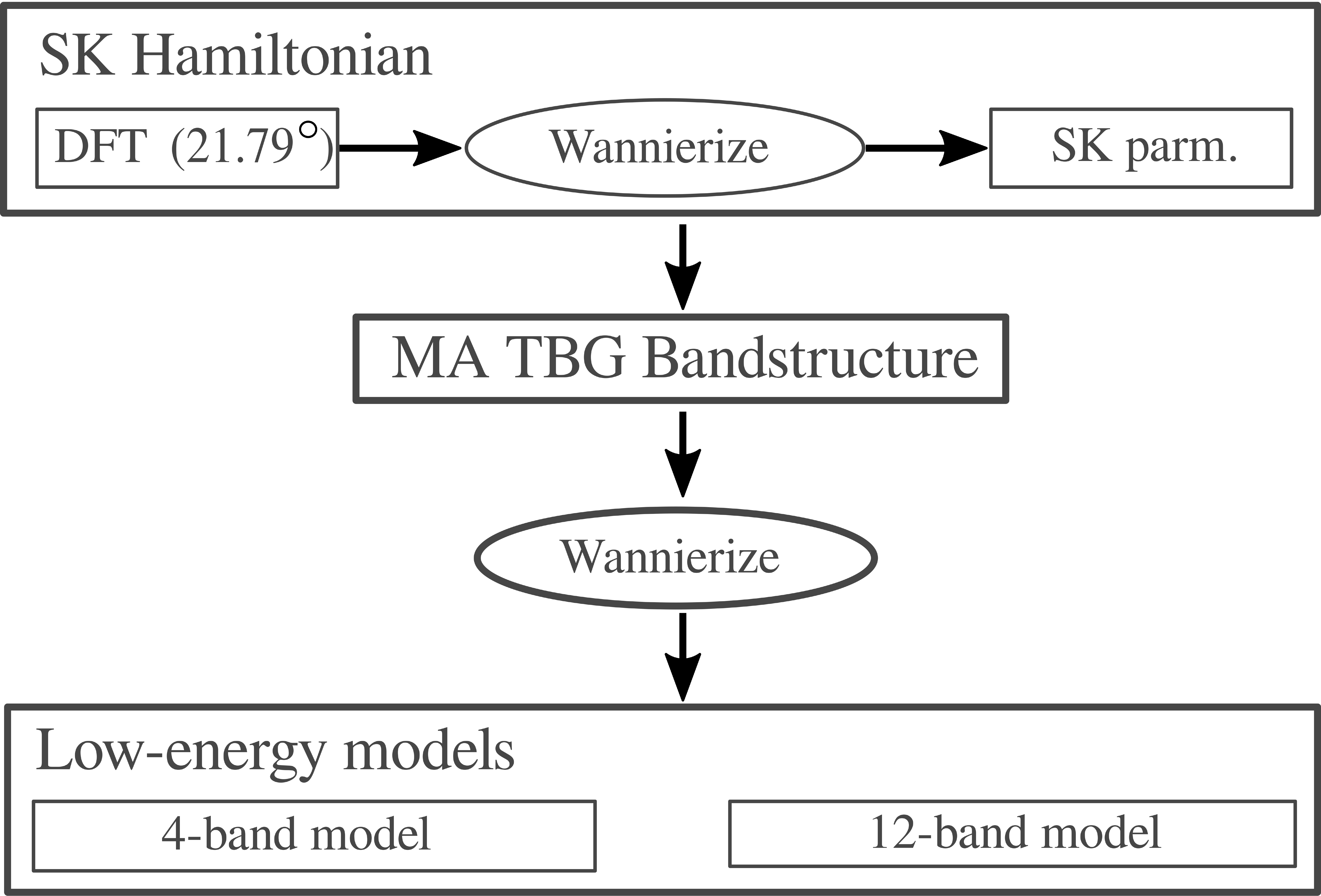}
\caption{The figure gives a workflow for creating TB Hamiltonians for TBG. The first line describes the process of creating the full energy range SK Hamiltonian using a $21.79^\circ$-TBG DFT calculation with subsequent Wannierization (see Sec.~\ref{sec:wannierization}) followed by an SK parameter fit. The resulting SK bandstructure for magic-angle TBG is further Wannierized (second line) to give the low-energy TB Hamiltonians (third line).} 
\label{fig:tbmodels_schema}
\end{figure}

In this section, we describe our workflow of constructing tight-binding models for TBG. We start from \textit{ab-initio} data at a large, commensurate twist angle ($\theta = 21.79^{\circ}$) to construct a TB model via a procedure known as Wannierization. The resulting hopping amplitudes are fitted using the SK analytic form. These fitted SK parameters can then be used to construct the TB model at the magic angle, thus allowing us to bypass the expensive step of Wannierizing \textit{ab-initio} data at the magic angle. Finally, having obtained the TB model at the magic angle, a second Wannierization step is used to obtain the low-energy (four-band and twelve-band) TB Hamiltonians. The full scheme is depicted in Fig.~\ref{fig:tbmodels_schema}. In the rest of the section, we give a detailed explanation of each step in the procedure.

\subsection{Wannierization procedure\label{sec:wannierization}}
We first discuss the Wannierization, the process of creating a TB Hamiltonian, which describes a desired manifold of given Bloch states. If the desired set of $J$ bands is \textit{isolated} from other bands at all points in the Brillouin zone, the procedure proceeds simply by finding the set of $J$ maximally-localized Wannier orbitals $|w_{n\RR}\rangle$ related to the original Bloch function via a Fourier and unitary transformation, 
\begin{eqnarray}
    |w_{n\RR}\rangle&=&
    \frac{1}{N_\mathrm{BZ}} \sum_{\kk} e^{-i\kk\RR}
    \sum_{m=1}^{J}U_{mn}^{\kk} |\psi_{m\kk}\rangle \nonumber\\
    &=&\frac{1}{N_\mathrm{BZ}}\sum_{\kk} e^{-i\kk\RR} |\phi_{n\kk}\rangle,
    \label{eq:Wannier_orbital_from_Bloch}
\end{eqnarray}
where the $\kk$ sum runs over $N_\mathrm{BZ}$ momenta in the Brillouin zone, $U_{mn}^\kk$ are unitary matrices, which mix the Bloch states at each $\kk$, and $\RR$ are the unit cell position vectors with the so-called home unit cell located at the origin ($\RR=0$). The $U_{mn}^\kk$ matrices are chosen such that the spread
\begin{equation}
    \Omega=\sum_{n=1}^{J}
    \left[
       \langle w_{n\bf{0}}|\hat{\rr}\cdot\hat{\rr}|w_{n\bf{0}}\rangle
       -
       \left|\langle w_{n\bf{0}}|\hat{\rr}|w_{n\bf{0}}\rangle\right|^2
    \right]
\end{equation}
is minimized. Here $\hat{\rr}$ is the position operator. The spread $\Omega$ can be decomposed into two positive-definite components
\begin{equation}
    \Omega=\Omega_I+\widetilde{\Omega},\label{eq:Omega_via_OmegaIOmega_tilde}
\end{equation}
where $\Omega_I$ is gauge-independent while $\widetilde{\Omega}$ is gauge-dependent. When the target bands are isolated, only the gauge-dependent part needs to be minimized since the gauge-independent part is invariant under unitary transformation. Technically, this minimization is achieved using a steepest descent algorithm, for which one needs to compute the gradient
\begin{equation}
    \mathcal{G}_{\kk}=\mathrm{d}\widetilde{\Omega}/\mathrm{d}W^{\kk},\label{eq:Gradient}
\end{equation}
where the antihermitian $\mathrm{d}W^{\kk}$ generates the infinitesimal gauge transformation as $U^\kk_{mn}=\delta_{mn}+\mathrm{d}W^{\kk}_{mn}$~\cite{Marzari1997,Pizzi2020}. At each iteration, the unitary transformation is updated according to
\begin{equation}
     U^{\kk}\rightarrow U^{\kk} \exp[\alpha \mathcal{G}_{\kk}]\label{eq:rotate_U}
\end{equation}
with step variable $\alpha\leq 1$.

In the case of \textit{entangled} bands~\cite{PhysRevB.65.035109}, i.e., when the target set of bands is not energetically separated from other bands in the entire Brillouin zone, the above procedure has to be augmented with a disentanglement step. This step ensures that at each $\kk$ point the smoothest set of $J$ Bloch wavefunctions $|\widetilde{\psi}_{n\kk} \rangle$ is selected from a larger set of size $\mathcal{J}_\kk \geq J$~\cite{Pizzi2020},
\begin{equation}
    |\widetilde{\psi}_{n\kk} \rangle=\sum_{m=1}^{\mathcal{J}_\kk} V_{mn}^{\kk} |\psi_{m\kk} \rangle .\label{eq:dis_matrices_action}
\end{equation}
Here, $V^\kk$ are $\mathcal{J}_\kk\times J$ semi-unitary matrices ($[V^{\kk}]^\dagger V^\kk=\openone$), chosen such that the gauge-invariant spread $\Omega_I$ computed on $\{|\widetilde{\psi}_{n\kk}\rangle\}$ is minimized. Technically, this results in an iterative procedure, where at each iteration the mismatch of Hilbert subspaces, $\mathrm{span}\{ |\widetilde{\psi}_{n\kk} \rangle\}$ and $\mathrm{span}\{ |\widetilde{\psi}_{n\kk'} \rangle\}$, where $\kk'$ runs over all momentum points nearest to $\kk$, is minimized for all $\kk$.

In this work, we use an additional restriction in the disentanglement procedure~\cite{PhysRevB.65.035109}: in Sec.~\ref{sec:extraction_hopping} we choose an energy range within which the original Bloch \mbox{manifold} must be exactly reproduced. This energy range is called the \textit{inner}, or \textit{frozen} window of the disentanglement procedure. In this case, the manifold of Bloch states within the frozen window is fixed and always contained in $\mathrm{span}\{ |\widetilde{\psi}_{n\kk} \rangle\}$.

Due to the non-convex nature of the optimizations, the initialization of the various iterative procedures plays a critical role. This initialization is achieved by choosing a set $\{|\omega_n\rangle\}$ of trial orbitals.
Given the trial orbitals, the initial value of the matrices $U^\kk$ and $V^\kk$ can then be defined in the following way. First, one expands the projection above back onto the Hilbert space of interest
\begin{equation}
    |\bar{\phi}_{n\kk}\rangle
    =\sum_{m=1}^{J \mathrm{\ or\ } \mathcal{J}_\kk}
   A_{mn\kk} |\psi_{m\kk}\rangle ,
\end{equation}
where
\begin{equation}
    A_{mn\kk}=\langle\psi_{m\kk}|\omega_{n}\rangle.\label{eq:Amn}
\end{equation}
Next, by orthonormalizing, we obtain a set of Bloch states, in the sense of $|\phi_{n\kk}\rangle$ or $|\widetilde{\psi}_{n\kk}\rangle$ from Eqs.~\eqref{eq:Wannier_orbital_from_Bloch} and~\eqref{eq:dis_matrices_action}, respectively:
\begin{eqnarray}
    |\phi_{n\kk}\rangle
    &=&
    \sum_{m=1}^{J} S^{-{\frac{1}{2}}}_{mn\kk}|\bar{\phi}_{m\kk}\rangle \\
    &=&
    \sum_{m=1}^{J \mathrm{\ or\ }  \mathcal{J}_\kk}
   (A_\kk S^{-{\frac{1}{2}}}_\kk)_{mn} |\psi_{m\kk}\rangle ,\label{eq:phi_initial}
\end{eqnarray}
where $S_{mn\kk}=\langle\bar{\phi}_{m\kk}|\bar{\phi}_{n\kk}\rangle=A^\dagger_\kk A_\kk$. The matrices $(A_\kk S^{-{\frac{1}{2}}}_\kk)_{mn}$ then serve for initialization of  $U^\kk_{mn}$ and $V^\kk_{mn}$ in the maximum-localization and disentanglement algorithm, respectively. When the trial orbitals are chosen appropriately, this initial guess for $U^{\kk}_{mn}$ is often used for TB model construction, i.e., without performing maximum localization. In Sec.~\ref{sec:extraction_hopping}, we make use of such a one-step procedure after obtaining a converged disentangled manifold.

After obtaining the $U^\kk$ and $V^\kk$ matrices, the TB Hamiltonian can be constructed by rotating the initial eigenvalues as
$U^\kk \hat{E}^{\kk} [U^{\kk}]^\dagger$
or $U^\kk V^\kk \hat{E}^{\kk} [V^{\kk}]^\dagger [U^{\kk}]^\dagger$
in the entangled-bands case. Here, $\hat{E}^{\kk}$ is the matrix with the energy eigenvalues on the diagonal, and zero entries otherwise. Then, the subsequent (inverse) Fourier transform to real space gives the desired set of hopping parameters $t_{n_1n_2}(\RR)$, such that the translationally invariant TB Hamiltonian in second quantized form reads
\begin{equation}
    \hat{H}=\sum_{n_1\,n_2}\sum_{\RR_1\RR_2} t_{n_1n_2} (\RR_2-\RR_1) \cdag_{n_1 \RR_1}\cc_{n_2\RR_2},
\end{equation}
where  $\hat{c}_{n\RR}$ $(\hat{c}^\dagger_{n\RR})$ are second quantized  operators, which create (annihilate) a Wannier orbital $|w_{n\RR}\rangle$ (see Eq.~\eqref{eq:Wannier_orbital_from_Bloch})  located at position $\rr_{n\RR}=\RR+\btau_n$ with $\btau_n$ being the position  of orbital $|w_{n\RR}\rangle$ within the unit cell.

\subsection{Crystal symmetry constraint\label{sec:site-symmetry}}
The Wannierization procedure defined above preserves neither crystal nor TR symmetries. For the procedure to respect these symmetries, we place a constraint on the matrices $U^\kk$ and $V^\kk$ during the maximal-localization and disentanglement procedures, respectively~\cite{Sakuma2013,Pizzi2020}. 

To simplify the notations, we restrict ourselves to working with the point group $G$ only, which is possible in TBG by choosing the unit cell appropriately. 
In the space of Bloch states the symmetry representation matrix $\tilde{d}^{g}_{mn} (\kk)$ for a point group element $g\in G$  is defined through
\begin{equation}
    \g|\psi_{n\kk}\rangle 
    = \sum_{m=1}^{J} \tilde{d}^{g}_{mn} (\kk) |\psi_{m, S_g\kk} \rangle
    , \label{eq:d_tilde_mn}
\end{equation}
where the action of 
$g$ in euclidean space is expressed as $g \rr=S_{g}\rr$. On the other hand, the corresponding Wannierized states $|\phi_{n\kk}\rangle$ have to obey a similar transformation rule
\begin{equation}
   \g |\phi_{n\kk}\rangle
   = \sum_{m=1}^{J} D^{g}_{mn} (\kk) |\phi_{m, S_g\kk} \rangle 
   ,\label{eq:D_irrep_mn} 
\end{equation}
with desired symmetry representation matrices $D^{g}_{mn} (\kk)$, which are block diagonal, and each block corresponds to a site-symmetry-induced irreducible representation determined from the chosen orbital configuration of the TB model. The exact definitions of $\tilde{d}^{g}_{mn} (\kk)$ and $D^{g}_{mn}(\kk)$ can be found in Ref.~\cite{Sakuma2013}. It can be shown
that for a point group operation $g_\kk$ in the little group $G_\kk$---the subgroup of $G$ which leaves a given $\kk$ unchanged---the following relationship holds~\cite{Sakuma2013} 
\begin{equation}
    U^{\kk}D^{g_{_\kk}} (\kk)=\tilde{d}^{\hskip 0.05cm g_{_\kk}} (\kk) U^{\kk}\label{eq:UD_eq_dU}
\end{equation}
for the matrices $U^\kk$ from Eq.~\eqref{eq:Wannier_orbital_from_Bloch} (exactly the same equation must hold for the $V^\kk$ matrices from Eq.~\eqref{eq:dis_matrices_action} in the entangled-bands case). Starting from the initial guess, this equation is solved iteratively to ensure that the symmetry condition holds. Note, that Eq.~\eqref{eq:UD_eq_dU} can only be satisfied if the irreducible representations of original and the targeted Wannierized Bloch manifolds~\cite{Sakuma2013} are compatible in the targeted energy window. 

In the subsequent steepest descent optimization of the maximum localization procedure, the gradient $\mathcal{G}_{\kk}$ in Eq.~\eqref{eq:Gradient} is replaced by
\begin{equation}
    \mathcal{G}^\mathrm{sym}_{\kk}=\frac{1}{|G_{\kk}|} \sum_{g_{\kk}\in G_\kk} D^{g_{_\kk}}(\kk) \mathcal{G}_{\kk} [D^{g_{_\kk}}(\kk)]^{\dagger},
\end{equation}
where $|G_{\kk}|$ is the order of the little group $G_\kk$,
in order to preserve the equality~\eqref{eq:UD_eq_dU} at every iteration.
One only needs to solve Eq.~\eqref{eq:UD_eq_dU} on an irreducible wedge of momentum points, while $U^\kk$  at other momentum points is constructed by applying the symmetry transformations. A similar approach is used at the disentanglement step~\cite{Sakuma2013,Pizzi2020}. 

\subsection{Time-reversal symmetry constraint\label{sec:trev_symmetry}}
In spinless systems with TR symmetry, a real basis of maximally-localized Wannier orbitals is guaranteed to exist. While Wannier90 code usually manages to find such real Wannier orbitals, this is generally not guaranteed. Particularly in TBG, this symmetry tends to be broken when constructing low-energy TB models. The TR-symmetric four-band low-energy model can be obtained by considering a set of complex TR-related pairs as trial orbitals~\cite{Kang2018}, and skipping the maximum-localization procedure. Either when using maximum localization or an odd number of such complex orbitals at some sites (as in our twelve-band model below), the TR symmetry can not be fixed explicitly by available software, and therefore can be broken in the resulting TB Hamiltonian. In this work, we implement a scheme, which allows for explicit TR symmetry constraint in both disentanglement and maximum-localization steps, involving only a minor modification of Wannier90 code, and requiring no additional input.

First, we fix the arbitrary phase of eigenvectors of the original band structure to satisfy the TR symmetry constraint. Neglecting the spin degrees of freedom, this is achieved by only taking eigenvectors in the irreducible wedge of $\kk$ points with respect to the TR operator and constructing the eigenvectors at $-\kk$ point via
\begin{equation}
    |\psi_{-\kk}\rangle=\mathcal{T}|\psi_{\kk}\rangle,
\end{equation}
where the TR operator $\mathcal{T}=\mathcal{K}$, the complex conjugation. If the unitary matrices $U^\kk$ and $V^\kk$ satisfy the condition 
\begin{equation}
B^\kk=[B^{-\kk}]^{*},
\end{equation}
where $B^\kk$ corresponds to either $U^\kk$ or $V^\kk$, the resulting Wannier orbitals will be real. This condition can be satisfied by the substitution
\begin{equation}
    B^{\kk}\rightarrow\frac{1}{2} (B^{\kk}+[B^{-\kk}]^{*}).\label{eq:TRsymmetrize_U_isolated}
\end{equation}
However, the operation in Eq.~\eqref{eq:TRsymmetrize_U_isolated} breaks unitarity (for $U^\kk$) or semi-unitarity (for $V^\kk$) of the matrices, which has to be restored with an ``orthonormalization'' subroutine available within the Wannier90 software. 

Interestingly, it is only necessary to make the replacement Eq.~\eqref{eq:TRsymmetrize_U_isolated} in the final iteration of the maximum localization procedure. In this way, the replacement does not significantly affect the spread of the resulting Wannier orbitals and also preserves the crystal symmetries, discussed in Sec.~\ref{sec:site-symmetry}, accurately. On the other hand, for the disentanglement procedure, it is necessary to apply the substitution at every iteration. Further details regarding the symmetrization are provided in the appendix.

\subsection{Slater-Koster parametrization\label{sec:SK_approach}}
The hopping amplitudes of a TB Hamiltonian are often well described by the SK analytic form~\cite{Slater1954}. The analytic structure depends only on the geometric configuration, character of basis orbitals, and a set of fitting parameters. This approach was successfully applied to TBG by considering carbon's $p_z$ orbitals only with the following analytic form for the hopping amplitudes in the corresponding TB model~\cite{Koshino2018,TramblyDeLaissardiere2012,DeTramblyLaissardiere2010,Goodwin2019_cRPA,Goodwin2019,Haddadi2020}:
\begin{eqnarray}
\tij (\RR_j-\RR_i)&=&
    t_\pi(\rij)\left[1-\left(\frac{\zij}{\rij}\right)^2\right]
    +t_\sigma (\rij) \left(\frac{\zij}{\rij}\right)^2, \nonumber \\
t_\pi(r)&=&
    t^0_\pi \exp\left[q_\pi (1-r/r_{cc})\right],\nonumber \\
t_\sigma (r)&=&
t^0_\sigma \exp\left[q_\sigma (1-r/d_{ab})\right],\label{eq:SK_hop}
\end{eqnarray}
where $\rij=|\rr_{ij}|=|\rr_{j\RR_j}-\rr_{i\RR_i}|$, $\zij$ is the $z$ component of $\rr_{ij}$, $r_\mathrm{cc}$ is the in-plane carbon-carbon bond length, and $d_{ab}$ is the interlayer distance. The SK parameters $t^0_\pi$, $t^0_\sigma$, $q_\pi$ and $q_\sigma$, are typically fixed to $-2.7$ eV, $0.48$ eV, $3.14$ and $7.43$, respectively.

Due to its analytic form, the SK approach can be directly applied to handle the atomic relaxations specific to TBG. In this work, we consider only out-of-plane corrugations~\cite{Uchida2014}, in which the distance between layers varies from $d_{aa}$ in AA stacked regions  to $d_{ab}$ in AB stacked regions. Here,  AA (AB) stacking refers to regions where the same (opposite) sublattice atoms of the two graphene layers align. We use a smooth interpolation between these values, such that at the atomic site $\rr$, we have an interlayer separation
\begin{eqnarray}
   d(\rr)&=&d_{0}+2d_{1}[\cos({\bf b}_1\rr)+\cos({\bf b}_2\rr)\label{eq:corrugation}\\
         &+&\cos(\left\{{\bf b}_1+{\bf b}_2\right\}\rr)],\nonumber
\end{eqnarray}
where $d_{0}=(d_{aa}+2d_{ab})/3$, $d_{1}=(d_{aa}-d_{ab})/9$ with $d_{aa}=3.6$ \AA~and $d_{ab}=3.35$ \AA~\cite{Uchida2014,Koshino2018}, see Fig.~\ref{fig:Wyckoff}.
Although the average  magnitude of the in-plane and out-of-plane relaxations are similar in magic-angle TBG, our model of out-of-plane-only lattice corrugations yields a bandstructure very similar to the fully DFT-relaxed crystal structure of Ref.~\cite{Cantele2020}. As the corrugations generated by Eq.~\eqref{eq:corrugation} preserve the symmetries, we use this model in all our calculations. A comparison of different crystal structures and their corresponding electronic bands is shown in the Appendix~\ref{sec:compareCorrugations}. 

\subsection{Extraction of hopping parameters from \textit{ab-initio} calculation for TBG\label{sec:extraction_hopping}}
Usually, the SK parameters defining the hopping amplitudes in Eq.~\eqref{eq:SK_hop} are fixed to the values given in Sec.~\ref{sec:SK_approach}. However, this places a clear restriction on the accuracy of the resulting model. Therefore, we propose to Wannierize \textit{ab-initio} data at larger twist angles and use the resulting hopping amplitudes to extract the SK parameters. We assume that the  parameters extracted at large twist angles are still a good approximation for smaller angles. This way, we do not have to perform the expensive Wannierization using \textit{ab-initio} data at the magic angle.

We start by rewriting the TB Hamiltonian to distinguish between the in-plane ($\|$) and out-of-plane ($\perp$) hopping terms,
\begin{equation}
    \begin{split}
    \hat{H}
    =
    \sum_{\substack{ij\\\RR_i\RR_j}} 
        \bigl[
            &t^{\|}_{ij}(\RR_j-\RR_i) \Theta (z'_{ij})
            \\
            &+ 
            t^{\perp}_{ij} (\RR_j-\RR_i) \Theta (-z'_{ij})
        \bigr]\, \cdag_{i\RR_i} \cc_{n_j\RR_j},
    \end{split}\label{eq:SK_in-plane_ou-plane}
\end{equation}
where $z'_{ij}=\frac{1}{2}d_{ab}-|z_{ij}|$ and $\Theta$ denotes the Heaviside step function. The argument of the Heaviside step function allows to distinguish in-plane hopping with finite $z$-component from purely out-of plane components in the case of corrugated layers.

The TB parameters can then be obtained from the \textit{ab-initio} calculations by Wannierization of the DFT Bloch manifold, as discussed in Sec.~\ref{sec:wannierization}. Following Eq.~\eqref{eq:SK_hop}, the hopping coefficients $t_{ij}$ depend on the displacement $\rr_{ij}$ connecting the corresponding sites. 

Since the vertical component of $\rr_{ij}$ is small for in-plane hoppings, especially at small distances due to the long-wavelength nature of atomic corrugations, one may assume that the corresponding in-plane hopping amplitudes are only of $\pi$-type and identical to those in single-layer graphene (SLG).
Furthermore, the weak van der Waals forces of the interlayer coupling should only have a minor effect on the in-plane Hamiltonian of each individual layer of TBG. Still, we use the standard SK parametrization for the in-plane TB Hamiltonian in TBG, and find that together with our new out-of-plane parameterization (below) it produces the right bandstructures with the crystal structure defined in Eq.~\eqref{eq:corrugation} and the DFT-relaxed crystal structure from Ref.~\cite{Cantele2020}  (see Appendix~\ref{sec:compareCorrugations} for more details)~\footnote{Direct wannierization leads to a different sign of the second-nearest neighbor interaction in comparison with the first nearest-neighbor one appearing in the SLG ab-initio TB model. This sign structure of the TB parameters is confirmed in other works on the Wannierization of the DFT bandstructures for the SLG~\cite{Jung2013,Jung2014}. This is in contrast to fitting the DFT bandstructure (e.g. in Ref.~\cite{Reich2002}), which produces a TB model with hopping amplitudes all being negative up to the third nearest-neighbor, which probably can be achieved with the maximum localization algorithm as well subjected, however, to another constraint.}

For the computation of the out-of-plane hopping amplitudes $t^{\perp}_{ij} ({\RR_j-\RR_i})$, we follow a different path: Since the number of non-equivalent matrix elements of this type is very large at small twist angles, our approach is to ($i$) perform the Wannierization at larger twist angles (smaller unit cell), ($ii$) get the list of all possible out-of-plane hopping amplitudes for this case, and ($iii$) fit this data to the SK analytic expression. We apply this scheme to TBG at the largest and the second-largest commensurate twist angles $\theta=21.79^\circ$ and $\theta=13.17^\circ$, respectively. Finally, the resulting analytic expression for  $t^{\perp}_{ij} ({\RR_j-\RR_i})$ is used to construct the full TB Hamiltonian for TBG at arbitrary commensurate twists.

\begin{figure}
\centering
\includegraphics[width=0.47\textwidth]{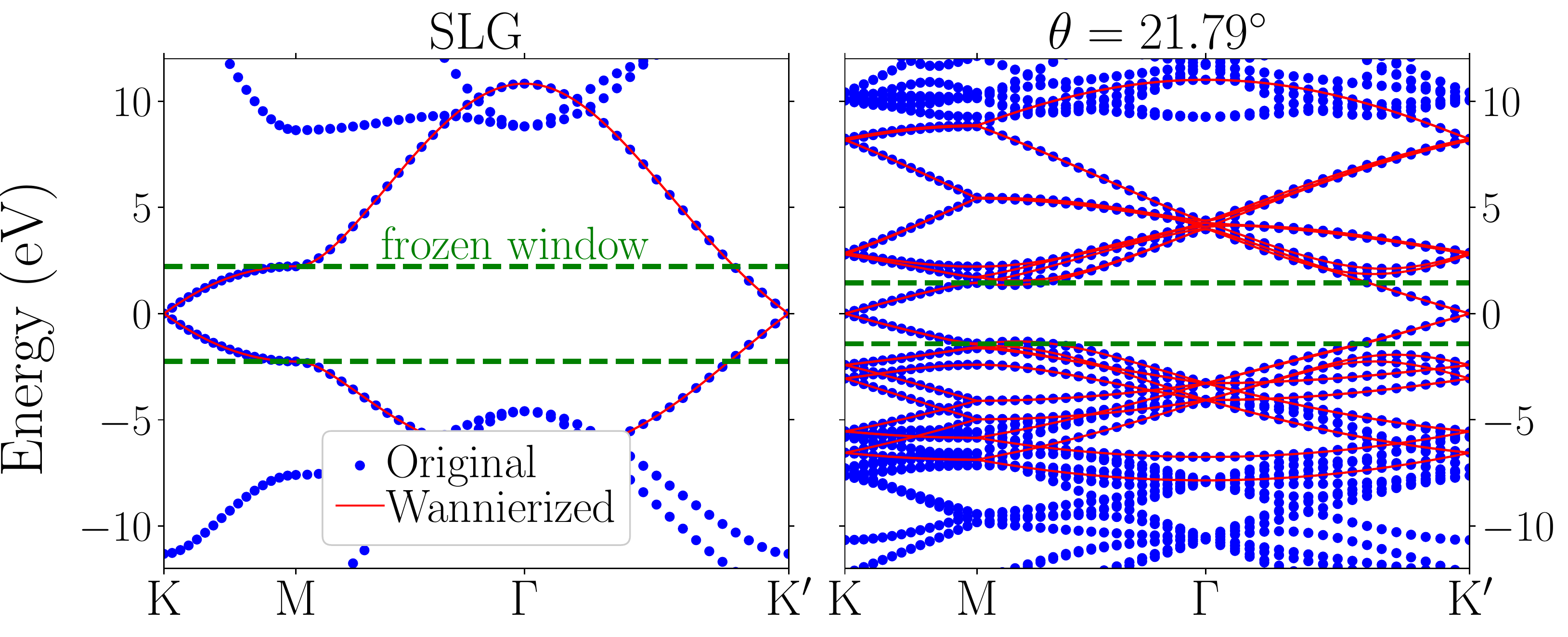}
\caption{Frozen windows (between dashed horizontal lines) for Wannierization of the DFT bandstructures for single layer and twisted bilayer graphene at $21.79^\circ$ respectively. The resulting Wannier-interpolated bandstructures (solid lines) and original DFT ones (dots) are also shown. The derived TB model for SLG was used in additional studies of Appendix~\ref{sec:compareCorrugations} and is placed here for a visual comparison.}
\label{fig:show_frozen_window}
\end{figure}

Since the maximum-localization procedure shifts, in general, the final Wannier-orbital centers from their original positions, we skip this step, i.e., the spread $\widetilde{\Omega}$ given in Eq.~\eqref{eq:Omega_via_OmegaIOmega_tilde} is not minimized, and the corresponding $U^\kk$ matrices are to be defined as described at the end of Sec.~\ref{sec:wannierization}. In addition, the disentanglement was done with the frozen window technique. The choice of the frozen window in both calculations is shown in Fig.~\ref{fig:show_frozen_window} with respect to the original DFT bandstructure. The Wannierized bandstructure is also shown in that figure.

\subsection{Low-energy models\label{sec:low-energy-models}}
The SK Hamiltonian described above is still large ($\sim 10^5\times10^5$ matrix per $\kk$ point), but can be exactly diagonalized as a non-interacting problem. Accounting for interactions, however, increases the computational costs exponentially. For many-body calculations, a smaller basis set is highly desirable. The most natural way to achieve such a smaller basis set is to perform a further Wannierization of the non-interacting TB Hamiltonian for TBG, such that the resulting model only contains bands close to the Fermi level. Unlike for the construction of the \textit{ab-initio} TB models above, we use a maximum localization procedure subjected to the crystal and time-reversal-symmetry constraints for  the construction of the low-energy models. The specific details on the choice of trial-orbital configurations are described in two sections below.

\subsubsection{Four-band Hamiltonian} Since the four flat bands of magic angle TBG are isolated from all other bands, the Wannierization to a four-band model is straight forward. It is known~\cite{Kang2018} that a good trial orbital set, $\lbrace |\omega_n\rangle \rbrace$ in Eq.~\eqref{eq:Amn}, is obtained by taking linear combinations of the Bloch wavefunctions at the $\Gamma$ point and truncating with Gaussians centered at the AB or BA site. In addition, we impose the crystal symmetry constraint on the Wannierization procedure as discussed in Sec.~\ref{sec:site-symmetry}.
In contrast to Ref.~\cite{Kang2018}, where complex orbitals are obtained, our target symmetry representation matrices are obtained by assuming that the Wannier functions transform as real $p_{x,y}$ atomic orbitals located at AB and BA sites of the TBG unit cell. Real orbitals are obtained by adding the constraint of time-reversal symmetry as described in Sec.~\ref{sec:trev_symmetry}. Following an earlier study in Ref.~\cite{Koshino2018}, we justify this choice by the correct irreducible representations of the Bloch bands at the $\Gamma$ and $K$ points in the Brillouin zone under the $D_3$ symmetry group (which we exclusively deal with in this article).

\subsubsection{Twelve-band Hamiltonian} Given the small (computed as $\sim25$meV) band gaps between the flat bands and the rest of the band structure, the inclusion of at least the next eight bands may be relevant for many-body physics. The problem, however, is that some of these additional bands are entangled, i.e., degenerate or crossing, with higher or lower bands.
Therefore, an additional disentanglement step is needed to construct a smooth manifold of Bloch states in momentum space for subsequent maximum localization. 

Reference~\cite{PhysRevResearch.1.033072} proposed a multistep projection scheme for constructing the trial set of orbitals. The main disadvantage of this scheme is that one has to work with a valley-projected bandstructure, which is generally not the result of \textit{ab-initio} calculations. Secondly, the number of resulting Wannier orbitals is strongly constrained. Two orbitals are needed for flat bands and three orbitals are needed for each lower and upper lowest-energy (non-flat) subset of bands. As a result, one can work with either $2,5$ or $8$ bands per valley.

The approach we shall take, on the other hand, is not subject to such restrictions. We simply choose a trial orbital set $\{|\omega_n\rangle\}$, which has a good overlap with the wavefunctions of the SK bandstructure close to the Fermi level, and delegate the search for the optimal Wannier orbitals to the Wannierization procedures described in Sec.~\ref{sec:wannierization} with crystal and time-reversal symmetry constraints, using the Wannier90 code. In particular, the trial orbitals in the twelve-band case are computed from the wavefunctions at $\Gamma$ in the following way. Consider the twelve bands closest to the Fermi level, labelling them $n=1, \ldots,12$ in order of increasing energy. Next, the bands with indices $n$ and $m=13-n$ are paired together. The trial orbitals $|\omega_n\rangle$ are then formed by combining the pairs $|\psi_{n\Gamma}\rangle$ and $|\psi_{m\Gamma}\rangle$. 
In addition, these trial orbitals are truncated by Gaussians centered along the center of rotation of the corresponding irreducible representation. A similar approach was recently taken in the four-band case~\cite{Kang2018}. Details on the symmetry configuration are discussed in Sec.~\ref{sec:result_low_energy_models}.

\section{Interactions}\label{sec:interactions}
Having obtained the non-interacting low-energy model, the final ingredient for a full many-body Hamiltonian is to consider the Coulomb interaction between electrons.
In general, this interaction is screened by the surrounding media, which in our case comprises all electrons that do not enter the low-energy Hamiltonian, and by an environmental dielectric function accounting for the substrate on which TBG is placed. A straightforward way to take into account this screening is by making the constrained random phase approximation (cRPA) which we describe below.

After computing the screening, one can then compute the Hubbard parameters in order to construct the many-body Hamiltonian. These parameters simply correspond to matrix elements of the cRPA-screened Coulomb interaction in the Wannier basis of the low-energy TB model. The workflow for calculating theses quantities is described in the subsequent sections and follows very closely  that of Ref.~\cite{Goodwin2019_cRPA}, with the exception that  symmetries are taken into account. In addition, we fit the resulting matrix elements with an analytic envelope function to better understand their dependence on distance.

\subsection{Screened Coulomb interaction}
The cRPA screening is described by the  polarizability function
\begin{equation}
    \Pi(\qq+\GG)=\frac{4}{\Omega} \sum_{\kk}\sum_{cv}^{}{}^{'}
    \frac{|\langle\psi_{v\kk}|e^{-i(\qq+\GG)\rr}|\psi_{c\kq}\rangle|^2}
    {\epsilon_{c\kq}-\epsilon_{v\kk}},\label{eq:polarizability}
\end{equation}
where $\qq$ is a vector in the first Brillouin Zone and $\GG$ is a reciprocal lattice vector. The prime in the second summation indicates that transitions between bands used in the construction of the low-energy Hamiltonian are excluded. The indices $c$ and  $v$ thus run over all conduction and valence bands, respectively, that are not part of the low-energy Hamiltonian. The cRPA dielectric function is then evaluated via
\begin{equation}
    \varepsilon(\qq+\GG)=\varepsilon_{\mathrm{env}}+v(\qq+\GG) \Pi(\qq+\GG),
\end{equation}
with $v(\qq+\GG) = {2\pi}/{|\qq+\GG|}$ the bare Coulomb interaction in two dimensions in atomic units and we choose a substrate-specific $\varepsilon_{\mathrm{env}} = 5$ to simulate  screening~\cite{Goodwin2019_cRPA}. The screened Coulomb interaction can then  be computed in real space by performing the Fourier transform:
\begin{equation}
    W(\rr)=\sum_{\GG}\int   \frac{d\qq}{(2\pi)^2} \frac{v(\qq+\GG)}{\varepsilon(\qq+\GG)} e^{-i\qq\rr}.\label{eq:FT_2d}
\end{equation}

The eigenvalues $\epsilon_{n\kk}$ entering the cRPA polarizability expression in Eq.~\eqref{eq:polarizability} are computed on a $7\times 7$ grid of $\kk$ points. In total, 8400 bands were used to ensure convergence. Calculation of the cRPA polarizability for the four-band model is straightforward. One simply needs to exclude four flat bands out of the sum in Eq.~\eqref{eq:polarizability}. The twelve-band model, on the other hand, is constructed with the disentanglement procedure, which forbids the direct use of Eq.~\eqref{eq:polarizability} in principle. However, due to the fact that the energy range of the twelve-band model spectrum is equal to that of the lowest twelve bands of the original Hamiltonian, we can assume that screening properties of both sets of bands should be similar. Moreover, the resulting disentanglement matrices of Eq.~\eqref{eq:dis_matrices_action} show that the dominant contribution of the final Wannier basis set indeed comes from these twelve bands. Therefore, we use Eq.~\eqref{eq:polarizability} as given ignoring the twelve bands closest to the Fermi level in the sum, which is straightforward since this set is isolated on the chosen $\kk$-point grid (the degenerate $K$ and $K'$ Brillouin-zone points are not present on this grid).

\subsection{Hubbard parameters\label{sec:Hubbard_parameters_formulas}}

The screened Coulomb interaction computed in the previous section is incorporated into the non-interacting low-energy TB Hamiltonian in the form of two-body interaction terms given by
\begin{equation}
    U_{n_3n_4}^{n_1 n_2}(\RR_2 - \RR_1) \hat{c}^{\dagger}_{n_1\RR_1} \hat{c}^{\dagger}_{n_2\RR_2} \hat{c}_{n_3\RR_2} \hat{c}_{n_4\RR_1},
    \label{eq:U_second_quant}
\end{equation}
where $U_{n_3n_4}^{n_1 n_2}(\RR_2 - \RR_1) $ is the two-body interaction matrix element. When $n_1 = n_4$ and $n_2 = n_3$, the terms become a density-density interaction and the corresponding parameters $U_{n_1 n_2} = U_{n_2 n_1}^{n_1 n_2}$ are also known as the Hubbard parameters.

Given the Wannier-orbital wave functions 
$w_{n\RR}(\rr) = \langle \rr|w_{n\RR}\rangle$ 
as well as the screened interaction obtained in the previous section, the Hubbard parameters can be computed as
\begin{eqnarray}
      U_{n_1n_2}(\RR_2-\RR_1)&=&\iint d\rr_1 d\rr_2\label{eq:HubU_normal} \\
      &\times&|w_{n_1\RR_1}(\rr_1)|^2 W(\rr_2-\rr_1) |w_{n_2\RR_2}(\rr_2)|^2. \nonumber
\end{eqnarray}
The calculation is done by decomposing the Wannier orbitals in the basis of $p_z$ orbitals $|\phi^z_{\btau_i+\RR'}\rangle$ at each carbon site $i$ of TBG
\begin{equation}
    |w_{n\RR}\rangle=\sum_{i\RR'} c_{ni}(\RR'-\RR) |\phi^z_{\btau_i+\RR'}\rangle.
\end{equation}
Here $c_{ni}(\RR'-\RR)$ are defined from the solution of the free SK-TB Hamiltonian $\langle\phi^z_{\btau_i+\RR'}|\psi_{m\kk}\rangle$, via the result of the maximum localization procedure
\begin{equation}
    c_{ni}(\RR'-\RR)=\frac{1}{N_{BZ}}\sum_{m\kk} U^{\kk}_{mn} e^{i\kk(\RR'-\RR)} \langle\phi^z_{\btau_i+\RR'}|\psi_{m\kk}\rangle,
\end{equation}
where $U^{\kk}_{mn}$, the transformation from Eq.~\eqref{eq:Wannier_orbital_from_Bloch}, always carries a bold momentum superscript, and should not be confused with the Hubbard parameter matrix without such a superscript. In the twelve-band case, the orbitals $|\psi_{m\kk}\rangle$ must be replaced by smoothed $|\widetilde{\psi}_{m\kk}\rangle$ from Eq.~\eqref{eq:dis_matrices_action}.

Following the approach of Refs.~\cite{Goodwin2019,Goodwin2019_cRPA}, we have assumed that carbon $p_z$ orbitals of different sites have zero overlap, such that the final expression for the Hubbard parameters is given by
\begin{eqnarray}
    U_{n_1n_2}(\RR_2-\RR_1)
    &=&\sum_{ij\RR'\RR''}W(\rr_{ij\RR'\RR''})\label{eq:Upzbasis}\\ 
    &\times& |c_{n_1i}(\RR'-\RR_1)|^2|c_{n_2j}(\RR''-\RR_2)|^2,\nonumber
\end{eqnarray}
where $\rr_{ij\RR'\RR''}=\RR''+\btau_j-\RR'-\btau_i$. 

The summation in Eq.~\eqref{eq:Upzbasis} needs to be cut at short distances due to the high computational costs. This leads to a slight error of the sum, and an associated small symmetry breaking of the matrix $U$. Moreover, the calculation of the full matrix  as given in Eq.~\eqref{eq:U_second_quant} is computationally more challenging. Therefore, we reduce the computation to matrix elements that are not related by symmetry (i.e., an irreducible wedge) and obtain the others via the application of symmetry transformations.
Symbolically, this operation can be represented as
\begin{equation}
    U_{\nn\RR}=\hat{O}_{\mm\RR'\leftarrow\nn\RR}U_{\mm\RR'},\label{eq:U_RedIrr}
\end{equation}
where $\nn,\mm$ are combined orbital indices (e.g., $\nn={n_1,n_2,n_3,n_4}$),  $\RR,\RR'$ are  lattice vectors for general [left hand side of Eq.~\eqref{eq:U_RedIrr}] and irreducible (right hand side) wedge respectively. $\hat{O}$ is a symmetry operation which connects these matrix elements.

\begin{widetext}
The transformation requires the full interaction matrix parameters defined by 
\begin{align}
   U^{n_1n_2}_{n_3n_4}(\RR_2-\RR_1)=
   &\langle w_{n_1\RR_1}w_{n_2\RR_2}  |\hat{W}|
    w_{n_3\RR_2} w_{n_4\RR_1} \rangle \nonumber\\
    =&\sum_{ij\RR'\RR''} c^*_{n_1i}(\RR'-\RR_1) c^*_{n_2j}(\RR''-\RR_2) W(\rr_{ij\RR'\RR''})  c_{n_3j}(\RR''-\RR_2) c_{n_4i}(\RR'-\RR_1).\label{eq:Uhub4pt}
\end{align}
When $n_1=n_4$ and $n_2=n_3$, the equation reduces to Eq.~\eqref{eq:Upzbasis}.

The operation in Eq.~\eqref{eq:U_RedIrr}, and the construction of the irreducible wedge is based on the knowledge of the symmetry transformations of the interaction parameters above, which in turn is based on the knowledge of site-symmetry representations of the Wannier orbitals, and thus, the point group representation matrices $D^g_{nm}$. The complete derivation for the symmetry transformations of $U$ is presented in Appendix~\ref{sec:symmetry_of_U}. Here, we give only the final expression,
\begin{align}
    \g U^{n_1n_2}_{n_3n_4}(\RR_2-\RR_1)
    &=\sum_{\substack{m_1m_2\\m_3m_4}}D^g_{n_1m_1}D^g_{n_2m_2} U^{m_1m_2}_{m_3m_4} (\TT_{n_2n_1}^{\widetilde{n}_2\widetilde{n}_1}) D^{\widetilde{g}}_{m_3n_3}D^{\widetilde{g}}_{m_4n_4} \label{eq:gU_MainText}\\
    &\times\delta_{\btau_{n_1},\btau_{n_4}}\delta_{\btau_{n_2},\btau_{n_3}}
    \delta_{\widetilde{\btau}_{n_1},\btau_{m_1}} 
    \delta_{\widetilde{\btau}_{n_2},\btau_{m_2}}
    \delta_{\btau_{m_3},\widetilde{\btau}_{n_3}}
    \delta_{\btau_{m_4},\widetilde{\btau}_{n_4}},\nonumber
\end{align}
where $\TT_{n_2n_1}^{\widetilde{n}_2\widetilde{n}_1}=S_{\widetilde{g}}(\RR_2-\RR_1)
    +S_{\widetilde{g}}(\btau_{n_2}-\btau_{n_1})
    -(\widetilde{\btau}_{n_2}-\widetilde{\btau}_{n_1})$, 
and tilde is related to the action of the inverse symmetry operation $\widetilde{g}=g^{-1}$ (see Appendix~\ref{sec:symmetry_of_U} for details). 
Note that Eq.~\eqref{eq:gU_MainText} can also be used to symmetrize an existing $U$ matrix, in a similar fashion as the symmetrization of TB models in Ref.~\cite{Gresch2018} (see Appendix~\ref{sec:symmetry_of_U}).

\end{widetext}

\section{Results\label{sec:results}}

\subsection{Extraction of TB parameters}

\begin{figure}
\centering
\includegraphics[width=0.48\textwidth]{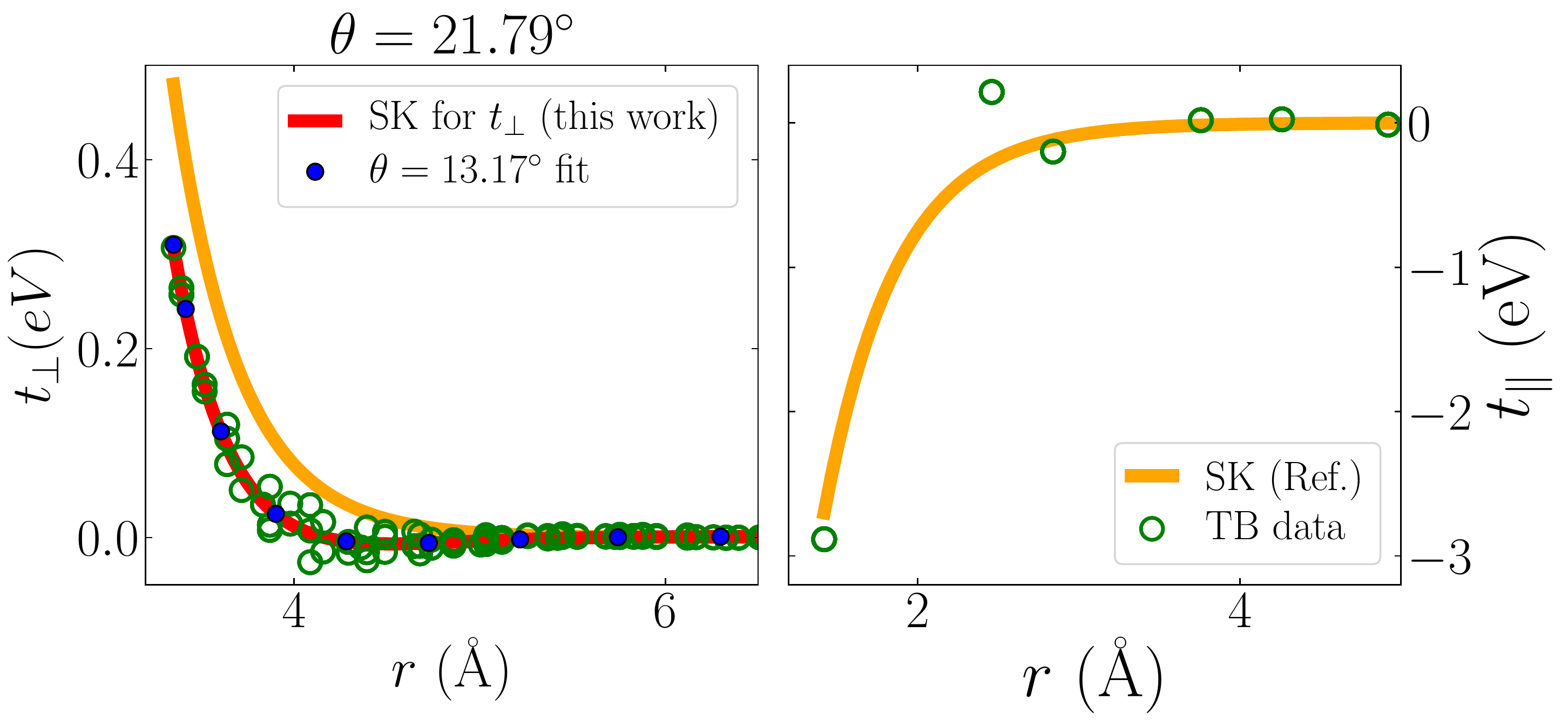} 
\caption{Data for the out-of plane (left) and in-plane (right) hopping amplitudes for $21.79^\circ$-TBG (empty circles) computed with the \textit{ab-initio} TB scheme.  Its SK parametric fit (red) together with the SK curve generated with commonly used parameters given in Sec.~\ref{sec:SK_approach}~\cite{Koshino2018,DeTramblyLaissardiere2010,TramblyDeLaissardiere2012,Goodwin2019,Goodwin2019_cRPA} (orange). The result of a similar fit for $13.17^\circ$-TBG for $t_\perp$ is given by filled circles on the left hand side. The good agreement between the SK fit at $21.79^\circ$ and the $13.17^\circ$ data supports our assumption that SK parameters are transferable between different angles.}
\label{fig:tb_parametrization}
\end{figure}

All Wannierization calculations were performed with Wannier90 code version 3.1.0~\cite{Pizzi2020}. The SIESTA code version 4.0.2~\cite{Garcia2020} was used for DFT calculations, for which a non-relativistic pseudopotential and the local density approximation (LDA) for the exchange-correlation functional was chosen. A $30\times30$ and $18\times18$ $\kk$ mesh was used in $21.79^\circ$ and $13.17^\circ$ TBG, respectively. In all cases, the so-called SZ (single-$\zeta$) basis of the SIESTA code was chosen.

Figure~\ref{fig:tb_parametrization} shows the in-plane and out-of-plane hopping amplitudes of $21.76^\circ$-TBG, as a function of distance. The values follow the SK analytical form well, albeit not with the standard parameters. In addition, we repeated the procedure using \textit{ab-initio} data for $13.17^\circ$-TBG. We found that the fitted SK parameters for this case are in good agreement with those computed for $21.76^\circ$-TBG, thus lending credibility for its subsequent use at the magic angle. Fitted Slater-Koster parameters for out-of-plane interactions can be found in Tab.~\ref{tab:SK_fit}. Notably large value of $t^0_\pi$ in comparison with its literature value of $-2.7$ eV is due to $1-[z_{ij}/r_{ij}]^2$ prefactor in Eq.~\eqref{eq:SK_hop}, which is small for closest out-of-plane interactions dominating the fit, and a larger value of the fitting parameter $t^0_\pi$ is required for a sensible range of the physical hopping parameters  $t_{ij}^{\perp}(\RR_j-\RR_i)$ [Eq.~\eqref{eq:SK_in-plane_ou-plane}].
\begin{table}
    \centering
    \begin{tabular}{c|cccc}
        & $t^0_\pi$ (eV)  & $t^0_\sigma$ (eV) & $q_\pi$ &  $q_\sigma$ \\
    \hline
   Ref. & -2.7      &   0.48     & 3.14    &  7.43       \\
   Fit  & -35.7    &   0.31     & 2.56    &  3.29       \\
    \end{tabular}
    \caption{Fitted Slater-Koster parameters for $t_{\perp}$. Referered values are used in Refs.~\cite{Koshino2018,DeTramblyLaissardiere2010,TramblyDeLaissardiere2012,Goodwin2019,Goodwin2019_cRPA} for describing both in- and out-of-plane hopping amplitudes.}
    \label{tab:SK_fit}
\end{table}

The slight variance of the data away from the SK-fitted line is explained by the fact that the actual electronic orbitals formed within the DFT framework are not strictly aligned along the vertical axes. An alternative scheme was proposed in Ref.~\cite{Fang2016} to include the dependence of the out-of-plane hopping amplitudes between atoms $i$  and $j$ on the in-plane $\pi$-bonding directions originating from these atoms. The reported small-angle TBG bandstructure supports a set of flat bands, which, however, is not as well isolated from other bands. This shortcoming has been attributed to neglected lattice relaxation effects. We leave the complete resolution of this question to future research. 

\subsection{Low-energy models\label{sec:result_low_energy_models}}

\begin{figure}
    \centering
    \includegraphics[width=0.48\textwidth]{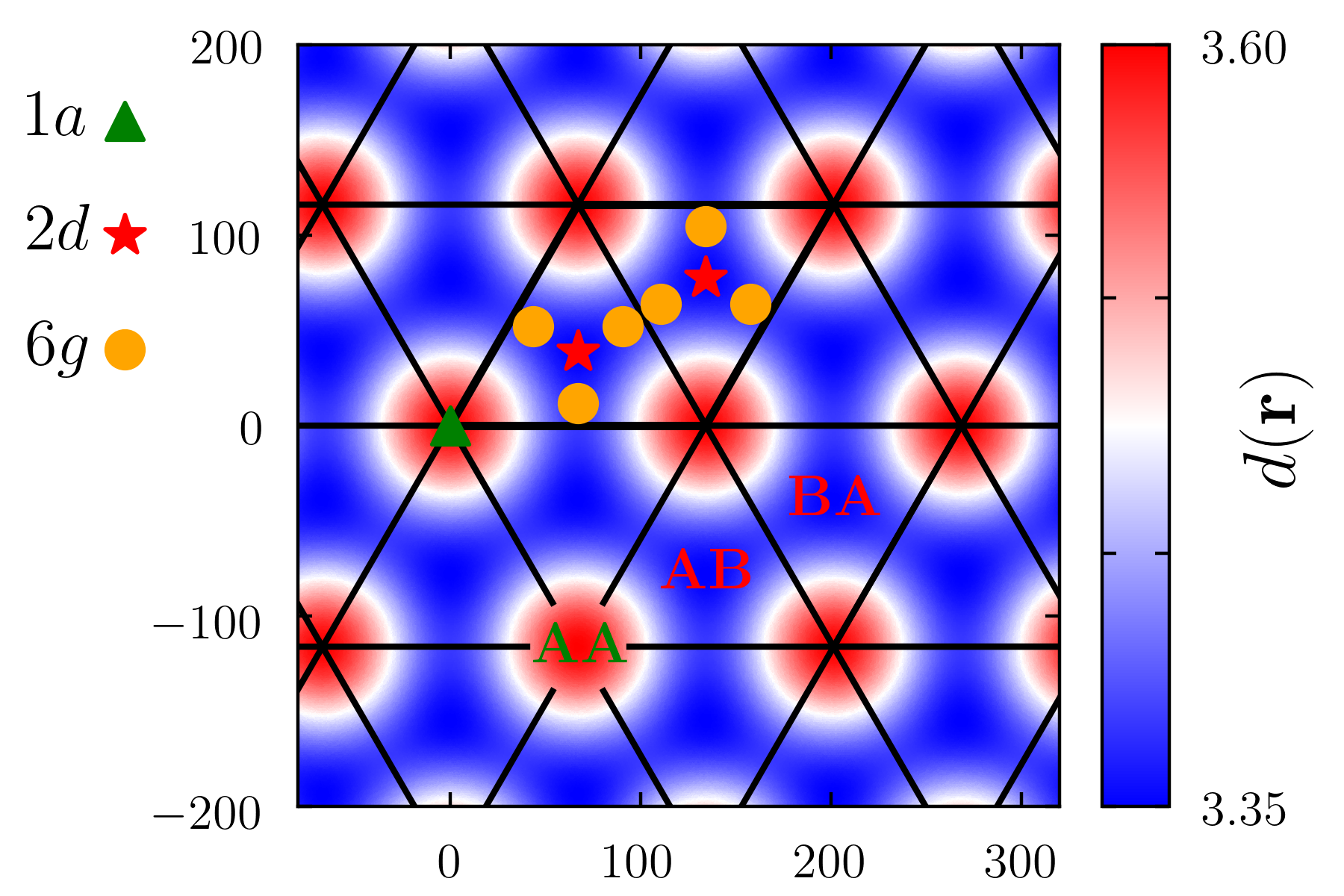}
    \caption{The corrugation structure of the magic-angle TBG together with all Wickoff positions used in this work as centers of trial orbitals. One of the $6g$ sites is taken as $(0.45,0.1)$ in lattice coordinates, the others are related by symmetry operations. The model of the out-of-plane corrugations seen on the plot is the one described by Eq.~\eqref{eq:corrugation}~\cite{Uchida2014,Koshino2018}, which gives comparable sizes of AA and AB domains of larger and smaller interlayer distance, respectively, encoded by the color. All distances are in \AA.} 
    \label{fig:Wyckoff}
\end{figure}

\begin{figure}
    \centering
    \includegraphics[width=0.48\textwidth]{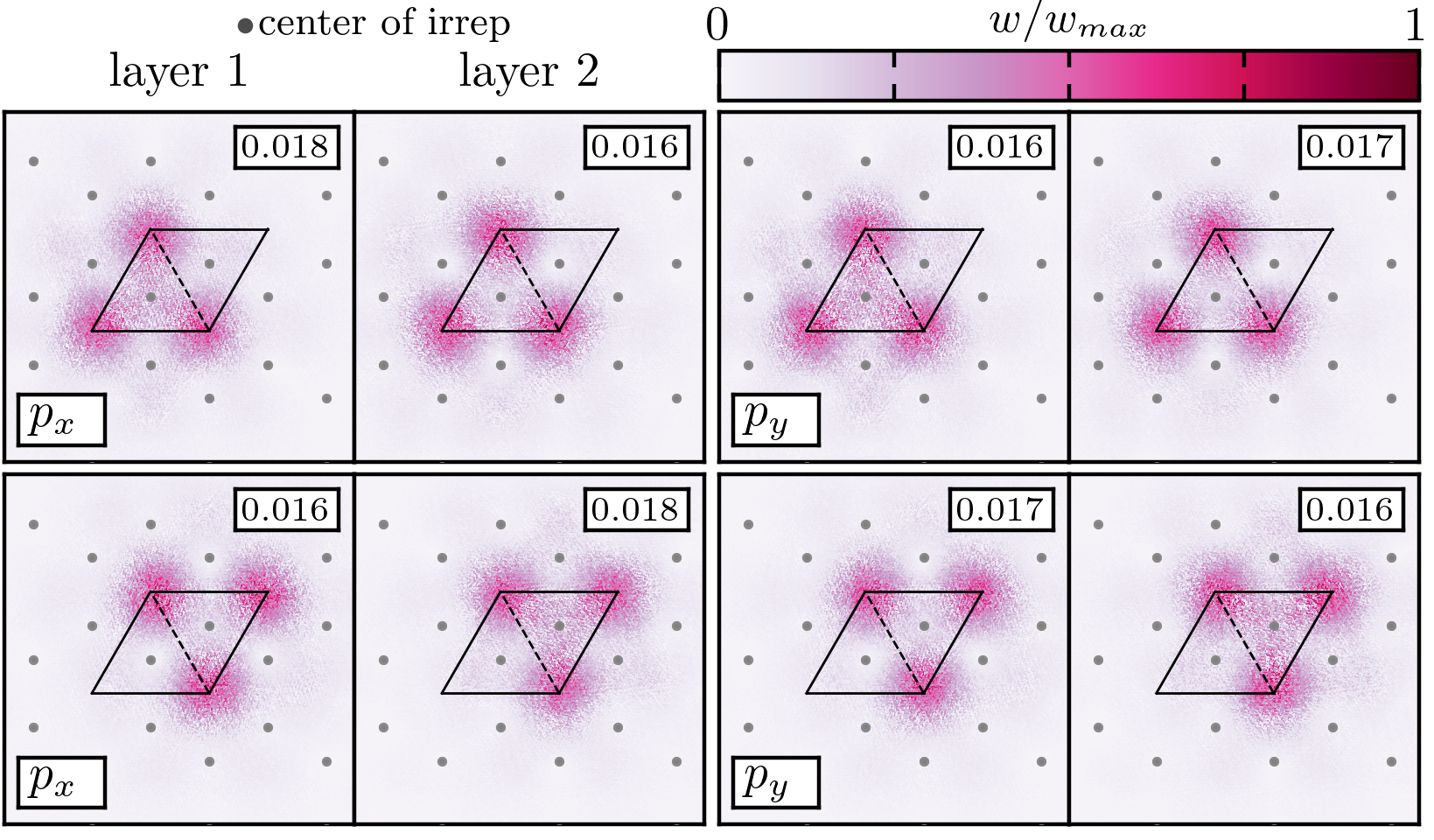}
    \caption{Wannier basis functions for four-band TB model. Color shows the weight of $p_z$ component relative to its maximal value specified at top-right of each subplot. Given Wannier functions transform according to atomic-like character shown at the bottom-left of each plot.} 
    \label{fig:wfs_4bands}
\end{figure}

\begin{figure}
    \centering
    \includegraphics[width=0.48\textwidth]{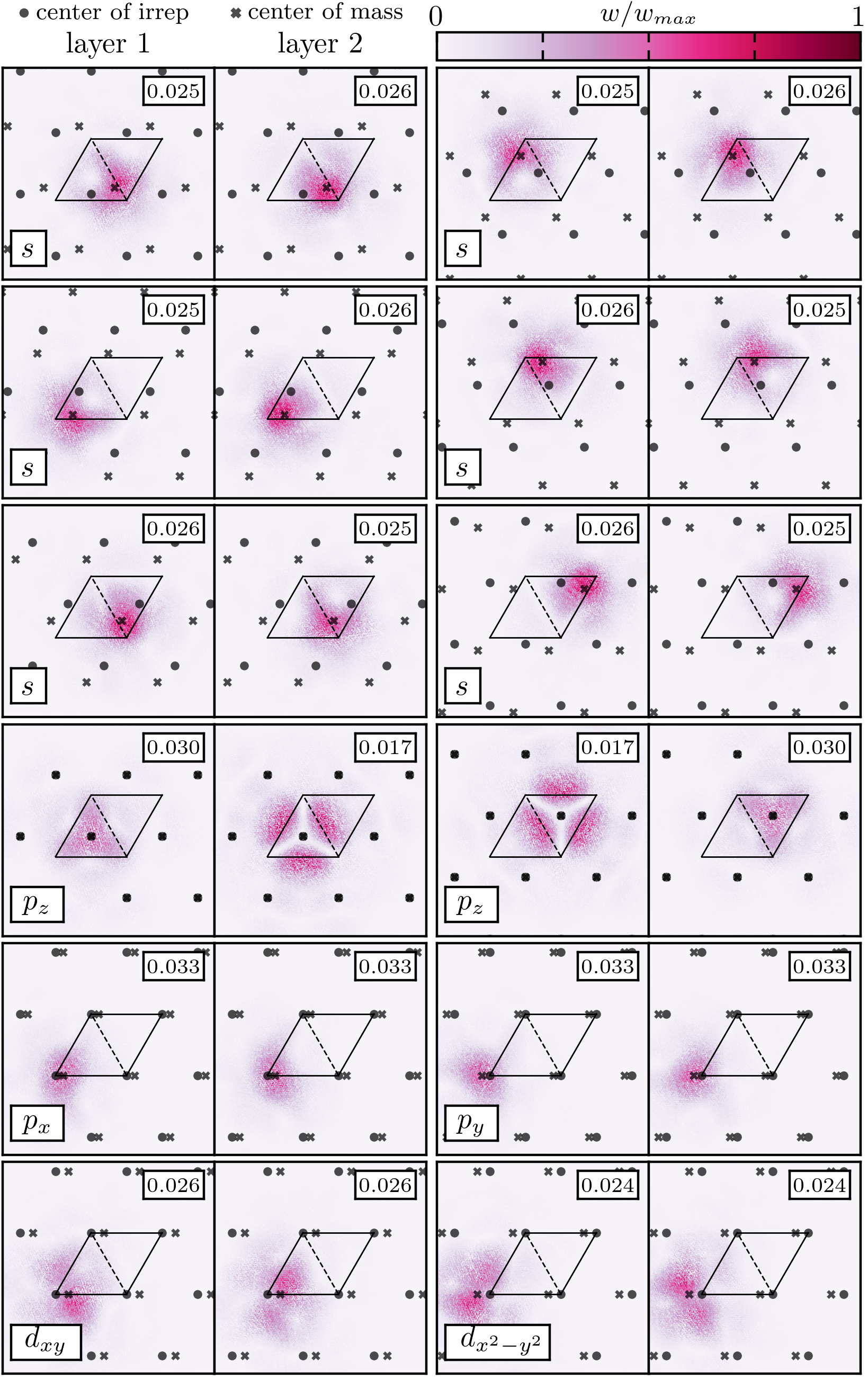}
    \caption{The same as Fig.~\ref{fig:wfs_4bands} but for twelve-band model. The difference is that centers of site-symmetry irreducible representations and the center of  mass of the orbitals do not generally match. In each plot only centers corresponding to the plotted orbital are shown.} 
    \label{fig:wannier_wf_12bands}
\end{figure}

\begin{figure}[t]
\centering
\includegraphics[width=0.48\textwidth]{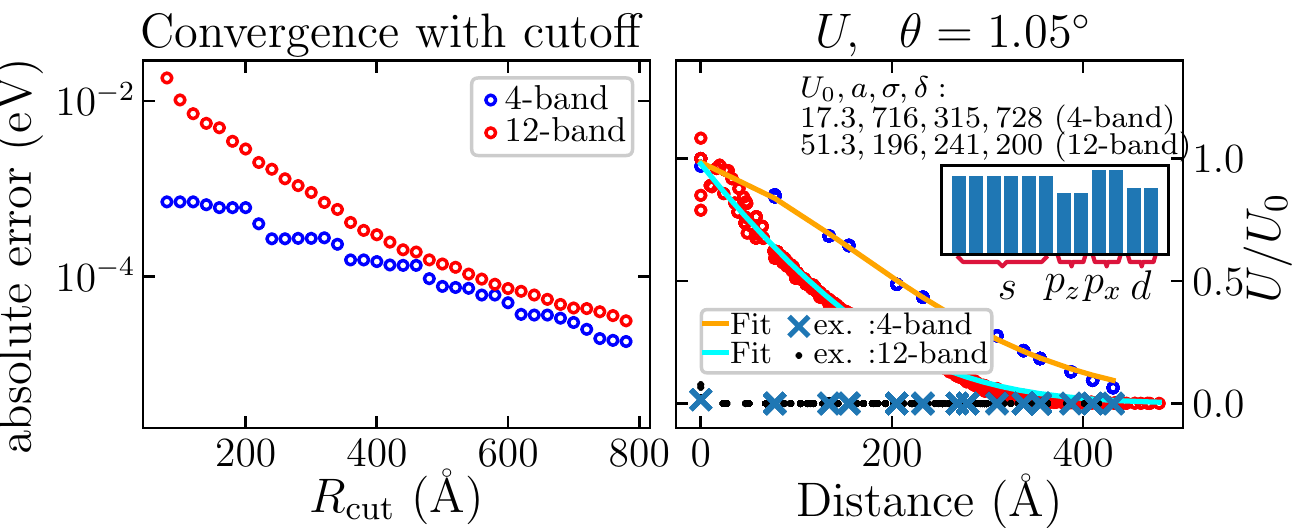}
\caption{(left) Averaged absolute error in a band energy per band and $\kk$-point computed with a low-energy Hamiltonian as a function of nearest-neighbor cut off $R_{cut}$. Note that the twelve-band model shows better convergence of flat bands (FB, green circles) than the four-band one. (right) Relative values of extended Hubbard parameters against the distance between centers of masses of orbitals. Fit to a soft Coulomb potential truncated by a Gaussian is given by the solid lines [Eq.~\eqref{eq:soft_coul_gauss}]. Fitted parameters are given as numbers in meV/\AA \ units. The inset shows the relative scale of the onsite intra-orbital Hubbard matrix elements for each orbital with value for $s$-orbitals given by $U_0$. The exchange parameters correspond to (ex.) label.}
\label{fig:bandtruct_error}
\end{figure}

In the construction of the four-band low-energy model, we used a $24\times24$ $\kk$-point mesh.
The trial orbital set was composed in a similar way as was done in Ref.~\cite{Kang2018} with a key difference that crystal and time-reversal symmetries were included in our calculations as discussed earlier. These symmetry constraints led to real-valued orbitals which transform as $p_{x,y}$, as opposed to the complex TR-related pairs in Ref.~\cite{Kang2018}. 
Another advantage of our scheme is that, in principle, one does not need the exactly symmetric trial orbital configuration, because this is fixed during the constrained maximum localization procedure. 

In the twelve-band case, the target symmetry configuration of the trial orbital set was chosen according to the one represented by atomic orbital symbols in Fig.~\ref{fig:wannier_wf_12bands}. It includes six $s$-like orbitals at the $6g$, two $p_z$-like at the $2d$, and a pair ($p_{x,y}$ and $d_{xy,x^2-y^2}$) of two-dimensional irreducible representations of $D_3$ at the $1a$ Wyckoff positions of space group $P321$~\cite{aroyo2006bilbao}. These Wyckoff positions are shown in Fig.~\ref{fig:Wyckoff}. This configuration gave a stable convergence of Eq.~\eqref{eq:UD_eq_dU} for several different $\kk$-point grids. We have chosen a $12\times12$ grid for the twelve-band TB model construction.

In both the four- and twelve-band cases, the Wannierization procedure results in exponentially localized Wannier functions as shown in Fig.~\ref{fig:wfs_4bands} and Fig.~\ref{fig:wannier_wf_12bands}, respectively. For the four-band model, the resulting orbital shape is similar to that obtained in Ref.~\cite{Kang2018}. On the other hand, the orbitals obtained for the twelve-band model have very diverse shape. Moreover, it can be seen from Fig.~\ref{fig:wannier_wf_12bands} that the center of mass and Wyckoff position do not necessarily match except when the orbitals are centered at AB and BA (honeycomb) sites. 

The spread of Wannier basis functions of the four- and twelve-band models can be quantified by computing the average absolute band energy error per band and $\kk$ point with respect to a hopping cutoff ($R_\mathrm{cut}$),
\begin{equation}
    \Delta(R_\mathrm{cut}) 
    = \frac{1}{N_\mathrm{bands} N_\mathrm{BZ}}\sum_{m,\kk} 
    |\epsilon_{m\kk} (R_\mathrm{cut}^\infty) -\epsilon_{m\kk}(R_\mathrm{cut}) |. 
\end{equation}
This is done by ignoring all hopping amplitudes above a certain threshold, i.e., when $|\rr_{n_2\RR_2}-\rr_{n_1\RR_1}|>R_\mathrm{cut}$ the hopping between sites $n_1\RR_1$ and $n_2\RR_2$ is ignored. $R_\mathrm{cut}^\infty$ corresponds to including all hopping terms. The result is shown in Fig.~\ref{fig:bandtruct_error}, and one can clearly see that the twelve-band model has shorter decay length, which can be attributed to the lower spread of orbitals. It is even possible to further reduce the spread of the Wannier orbitals in the twelve-band case by switching off the crystal symmetry constraint and using a frozen window in the disentanglement step. The result of the corresponding calculation is shown in Appendix~\ref{sec:non-symetric_12band}.

\begin{figure}
\centering
\includegraphics[width=0.48\textwidth]{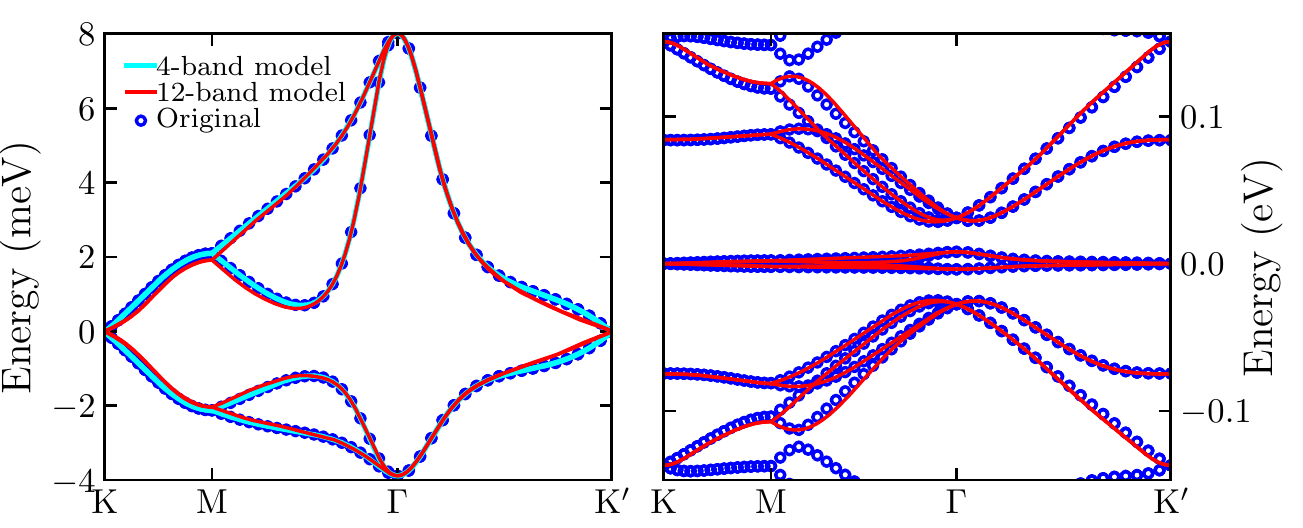}
\caption{Comparison between the original band structure of SK approach and its Wannier interpolation with the twelve-band model.}
\label{fig:bandstructure}
\end{figure}

Finally, we show the bandstructures of the resulting low-energy models in Fig.~\ref{fig:bandstructure}. There is a slight deviation of the twelve-band-model bandstructure around the Dirac points, which does not happen in the four-band case. This is due to the disentanglement step, which mixes bands outside the energy window of interest. This deviation and mixing is, however, very small and can be safely neglected.

\subsection{Hubbard parameters}
Hubbard parameters for both models are shown in Fig.~\ref{fig:bandtruct_error} (right). Fewer bands are included in the cRPA polarizability summation of Eq.~\eqref{eq:polarizability} of the twelve-band model in comparison to four-band case, which results in poorer screening. The orbitals are also more compact such that the Hubbard parameters have larger values at short range but decay faster with distance. For the four- and twelve-band models, the interactions become negligible at distances larger than $300$ \AA~and $500$~\AA, respectively, to be compared to the lattice constant of magic-angle TBG, $a\approx 134$ \AA. 

It was shown that the Ohno potential~\cite{Ohno1964} reasonably describes the dependence of Hubbard parameter values on the inter-orbital distance in TBG~\cite{Goodwin2019,Goodwin2019_cRPA}. However, a better fit was achieved in our calculations by considering a soft Coulomb interaction subjected to a Gaussian cut off,
\begin{equation}
    U^{\text{fit}}(r)=U_0\frac{a e^{-r^2/\sigma^2}}{|r+\delta|},\label{eq:soft_coul_gauss}
\end{equation}
where $U_0$ is the (fixed) on-site Hubbard interaction, while $a$, $\sigma$ and $\delta$ are fitting parameters. The fitted values for $U_0,a,\sigma,\delta$ are $14.4$ meV, $469$, $300$ \AA, $465$ \AA~and $57.7$ meV, $165$, $244$ \AA, $164$ \AA~for four- and twelve-band model respectively. The quality of this fit is demonstrated in Fig.~\ref{fig:bandtruct_error} (right). 

Finally, our exchange parameters, which correspond to taking $n_1=n_3$ and $n_2=n_4$ in Eq.~\eqref{eq:Uhub4pt} do not show any long-range behaviour, as seen from the Fig~\ref{fig:bandtruct_error}. This behaviour is different from the one found in Ref.~\cite{Cao2021}, and can be due to several reasons. First, we perform the maximum localisation procedure for obtaining the low-energy model. Second, we use full $D_3$ symmetry group instead of $C_{3z}$ one in the reference. Third, our RPA model of screening may behave differently than the gate screening used, most likely, in Ref.~\cite{Cao2021}. And last, we have used quite comprehensive tight-binding model, while the continuum model was employed in the reference above.

\section{Conclusions\label{sec:conclusion}}
In this work, we explored a new microscopic approach for obtaining an interacting low-energy tight-binding description of TBG. The two standard microscopic approaches begin either (i) by assuming that the TB hopping amplitudes are described by the SK form with a heuristic choice for the SK parameters or (ii) by performing a full Wannierization of \textit{ab-initio} DFT data at the magic angle. The former is computationally light but of limited accuracy, while the latter is computationally expensive despite being highly accurate. Our approach takes the middle ground by making a compromise between accuracy and computational cost.

The main idea is to perform Wannierization of \textit{ab-initio} data not at the magic angle, but at a larger twist angle, thus significantly reducing the computational requirements. The resulting TB model can then be used to extract the SK parameters which allows one to construct a TB model at arbitrary angles.  The bandstructure at the first magic angle was further Wannierized to give fully symmetric four- and twelve-band low-energy models. The low-energy Hamiltonians together with their corresponding Hubbard parameters are available at the open repository~\cite{HubbardTBG}. 

The full workflow presented in this article attempts to make a fair compromise between accuracy and computational cost. The various techniques and tricks used here are easily applicable and adaptable to any twisted multi-layer system, where the large unit cells present a computational difficulty.

During the course of this work we encountered two questions that might be worth studying in the future. First, the Wannierized \textit{ab-initio} data does not exactly follow the SK curves for interplane hopping amplitudes. This is most likely due to the assumption of rigid $p_z$ orbitals. Therefore, a potential improvement of the numerical precision of our scheme can be achieved by taking into account such deviations. Secondly, we found that a more compact twelve-band model can be obtained using the so-called frozen window technique when the crystal symmetry constraint is given up. The crystal symmetry constraint within the frozen window calculation is, however, not yet implemented in the Wannier90 code. Implementation of this functionality would allow to study  if such scheme would give a better result in comparison to the symmetric twelve-band model presented in the main text. This comparison is left for future research, when such implementation will be available.

\section*{Acknowledgement}
This work was initiated and motivated by Alexey Soluyanov, who sadly passed away in October 2019.

This research was supported by the NCCR MARVEL, funded by the Swiss National Science Foundation. K.C. is supported by the European Unions Horizon 2020 research and innovation program (ERC-StG-Neupert-757867-PARATOP). We thank Zachary Goodwin and Stepan Tsirkin for useful conversation and sharing the data. Finally, we thank Quansheng Wu and Giovanni Cantele for sharing their atomic structure data.

\bibliographystyle{apsrev4-2}
\bibliography{main}

\appendix

\begin{widetext}

\section{Symmetry transformations of Hubbard parameters\label{sec:symmetry_of_U}}
The transformation of the  Hubbard interaction presented in the main text  takes the following form:
\begin{eqnarray}
    \g U^{n_1n_2}_{n_3n_4}(\RR_2-\RR_1)&=&
    \langle w_{n_1\RR_1}w_{n_2\RR_2} |\g \hat{W} \g^{-1}| w_{n_3\RR_2} w_{n_4\RR_1}\rangle\\
    &=&
    \iint d\rr_1d\rr_2
    w^{*}_{n_1\RR_1}(\rr_1) w^{*}_{n_2\RR_2} (\rr_2) \g W(\rr_2-\rr_1) \g^{-1} w_{n_3\RR_2} (\rr_2) w_{n_4\RR_1}(\rr_1).\nonumber
\end{eqnarray}
The real-space integrals above must be computed over the entire space. For this purpose, we insert the complete set of basis orbitals from each side of $W$, which yields
\begin{eqnarray}
     \g U^{n_1n_2}_{n_3n_4}(\RR_2-\RR_1)
     &=&
    \sum_{\substack{m_1m_2m_3m_4\\ \TT_1\TT_2\TT_3\TT_4}}
    \langle w_{n_1\RR_1}w_{n_2\RR_2}|\g|w_{m_2\TT_2}w_{m_1\TT_1}\rangle
    \langle w_{m_1\TT_1}w_{m_2\TT_2} |\hat{W}| w_{m_3\TT_3}w_{m_4\TT_4}\rangle\\
    &\times&
    \langle w_{m_4\TT_4}w_{m_3\TT_3}|\g^{-1}|w_{n_3\RR_2} w_{n_4\RR_1}\rangle.\nonumber
\end{eqnarray}
Next, we introduce $g_1$ and $g_2$, \textit{equivalent} symmetry operations from the point group, acting on coordinates $\rr_1$ and $\rr_2$ correspondingly, such that we can further simplify
\begin{eqnarray}
    \g U^{n_1n_2}_{n_3n_4}(\RR_2-\RR_1)&=&
    \sum_{\substack{m_1m_2m_3m_4\\ \TT_1\TT_2\TT_3\TT_4}}
    \langle w_{n_1\RR_1}|\g_1|w_{m_1\TT_1}\rangle
     \langle w_{n_2\RR_2}|\g_2|w_{m_2\TT_2}\rangle
    U^{m_1m_2}_{m_3m_4}(\TT_1\TT_2\TT_3\TT_4)\\ 
    &\times&
    \langle w_{m_3\TT_3}|\g_2^{-1}|w_{n_3\RR_2}\rangle
    \langle w_{m_4\TT_4}|\g_1^{-1}|w_{n_4\RR_1}\rangle.\nonumber
\end{eqnarray}
If the underlying Wannier basis obeys known site symmetry representations, one is able to construct the representation matrices $D^g_{nm}$ and express matrix elements of the symmetry operations above accordingly:
\begin{equation}
        \langle w_{n\RR}|\g|w_{m\TT}\rangle = D^g_{nm}\delta_{\RR+\btau_n,\ybar{\TT}+\ybar{\btau}_m}, \label{eq:repmat}
\end{equation}
where $\btau$ is the center of Wannier orbital and $\xbar{\TT}+\xbar{\btau}$ gives the coordinates of transformed Wannier center by an operation $g$, such that $\xbar{\btau}$ is a Wannier center vector within the unit cell. Plugging this form into the equation above results in:
\begin{eqnarray*}
    \g U^{n_1n_2}_{n_3n_4}(\RR_2-\RR_1)&=&
    \sum_{\substack{m_1m_2m_3m_4\\ \TT_1\TT_2\TT_3\TT_4}}
    D^g_{n_1m_1}D^g_{n_2m_2}
    U^{m_1m_2}_{m_3m_4}(\TT_1\TT_2\TT_3\TT_4) 
    D^{\widetilde{g}}_{m_3n_3}  D^{\widetilde{g}}_{m_4n_4}\\
    &\times&
    \delta_{\RR_1+\btau_{n_1},\ybar{\TT}_1+\ybar{\btau}_{m_1}}
    \delta_{\RR_2+\btau_{n_2},\ybar{\TT}_2+\ybar{\btau}_{m_2}}
    \delta_{\TT_3+\btau_{m_3},\widetilde{\RR}_2+\widetilde{\btau}_{n_3}}
    \delta_{\TT_4+\btau_{m_4},\widetilde{\RR}_1+\widetilde{\btau}_{n_4}},
\end{eqnarray*}
where $\widetilde{\RR}+\widetilde{\btau}$ is transformed Wannier center by an inverse symmetry operation $\widetilde{g}=g^{-1}$. Any $\delta_{\RR+\btau_n,\ybar{\TT}+\ybar{\btau}_m}$ can be written as $\delta_{\widetilde{\RR}+\widetilde{\btau}_n,\TT+\btau_m}$. It follows that the combination of delta-symbols in the equation above transforms into:
\begin{eqnarray}
    &&\delta_{\TT_1+\btau_{m_1},\widetilde{\RR}_1+\widetilde{\btau}_{n_1}}
    \delta_{\TT_2+\btau_{m_2},\widetilde{\RR}_2+\widetilde{\btau}_{n_2}}
    \delta_{\TT_3+\btau_{m_3},\widetilde{\RR}_2+\widetilde{\btau}_{n_3}}
    \delta_{\TT_4+\btau_{m_4},\widetilde{\RR}_1+\widetilde{\btau}_{n_4}}\nonumber\\
    &\ &=
    \delta_{\TT_1,\widetilde{\RR}_1}
    \delta_{\TT_3,\widetilde{\RR}_2}
    \delta_{\TT_2,\widetilde{\RR}_2}
    \delta_{\TT_4,\widetilde{\RR}_1}
    \delta{\btau_{m_1},\widetilde{\btau}_{n_1}}
    \delta{\btau_{m_2},\widetilde{\btau}_{n_2}}
    \delta{\btau_{m_3},\widetilde{\btau}_{n_3}}
    \delta{\btau_{m_4},\widetilde{\btau}_{n_4}}\label{eq:delta_symbols1}.
\end{eqnarray}
Eq.~\eqref{eq:delta_symbols1} suggests, that the resulting translation vectors $\TT_1$ and $\TT_2$  after summing up the delta-symbols will be fixed as the following expressions:
\begin{eqnarray}
    \TT_1
    &=&\widetilde{\RR}_1+\widetilde{\btau}_{n_1}-\btau_{m_1}\\
    \TT_2
    &=&\widetilde{\RR}_2+\widetilde{\btau}_{n_2}-\btau_{m_2}
\end{eqnarray}
The $U$ depends on the difference of vectors above, and it could be computed via the following expression:
\begin{equation}
    \TT_2-\TT_1 \equiv \TT_{n_2n_1}^{\widetilde{n}_2\widetilde{n}_1}
    =S_{\widetilde{g}}(\RR_2-\RR_1)+S_{\widetilde{g}}(\btau_{n_2}-\btau_{n_1})
    -(\widetilde{\btau}_{n_2}-\widetilde{\btau}_{n_1}),
\end{equation}
where $S_{g}$ is the rotation matrix representing the point group operation $g$. Collecting all statements:
\begin{eqnarray}
    \g U^{n_1n_2}_{n_3n_4}(\RR_2-\RR_1)
    &=&
    \sum_{m_1m_2m_3m_4}
    D^g_{n_1m_1} D^g_{n_2m_2}U^{m_1m_2}_{m_3m_4}
    (\TT_{n_2n_1}^{\widetilde{n}_2\widetilde{n}_1})
    D^{\widetilde{g}}_{m_3n_3}D^{\widetilde{g}}_{m_4n_4}\\
    &\times&
    \delta_{\widetilde{\btau}_{n_1},\btau_{m_1}} 
    \delta_{\widetilde{\btau}_{n_2},\btau_{m_2}}
    \delta_{\btau_{m_3},\widetilde{\btau}_{n_3}}
    \delta_{\btau_{m_4},\widetilde{\btau}_{n_4}}.\nonumber
\end{eqnarray}
If we, use more strict constraint such that centers of orbitals $n_1,n_2$ coincide with ones of $n_4,n_3$ correspondingly, the transformation of the matrix $U$ will still be consistent:
\begin{eqnarray}
    \g U^{n_1n_2}_{n_3n_4}(\RR_2-\RR_1)
    &=&
    \sum_{m_1m_2m_3m_4}
    D^g_{n_1m_1} D^g_{n_2m_2} 
    U^{m_1m_2}_{m_3m_4}
    (\TT_{n_2n_1}^{\widetilde{n}_2\widetilde{n}_1}) 
    D^{\widetilde{g}}_{m_3n_3}D^{\widetilde{g}}_{m_4n_4}\\
    &\times&
    \delta_{\btau_{n_1},\btau_{n_4}}\delta_{\btau_{n_2},\btau_{n_3}}
    \delta_{\widetilde{\btau}_{n_1},\btau_{m_1}} 
    \delta_{\widetilde{\btau}_{n_2},\btau_{m_2}}
    \delta_{\btau_{m_3},\widetilde{\btau}_{n_3}}
    \delta_{\btau_{m_4},\widetilde{\btau}_{n_4}},\nonumber
\end{eqnarray}
meaning that such restricted $gU^{n_1n_2}_{n_3n_4}$ still depends on the same type of $U^{m_1m_2}_{m_3m_4}$ matrix elements in the sum, i.e., ones with $m_1,m_2$ and $m_4,m_3$ corresponding orbital centers matching, and these matrix elements contain the conventional Hubbard parameters at $n_1=n_4$ and $n_2=n_3$.

Finally, one could imagine a symmetrization of an existing $U$, if due to some reasons it does not obey the crystal symmetries, basically applying the same idea as for tight-binding models in Ref.~\cite{Gresch2018}:
\begin{equation}
    \mathbb{U}^{n_1n_2}_{n_3n_4}(\RR_2-\RR_1)=\frac{1}{N_g}\sum_{g\in G} \g  U^{n_1n_2}_{n_3n_4}(\RR_2-\RR_1).
\end{equation}
Validity of such expression was checked by applying this operation twice to originally non-symmetric U, the result of second symmetrization was the same as the first one.

\end{widetext}

\clearpage

\section{Non-symmetric twelve-band TB model.\label{sec:non-symetric_12band}}
\begin{figure}
\centering
\includegraphics[width=0.48\textwidth]{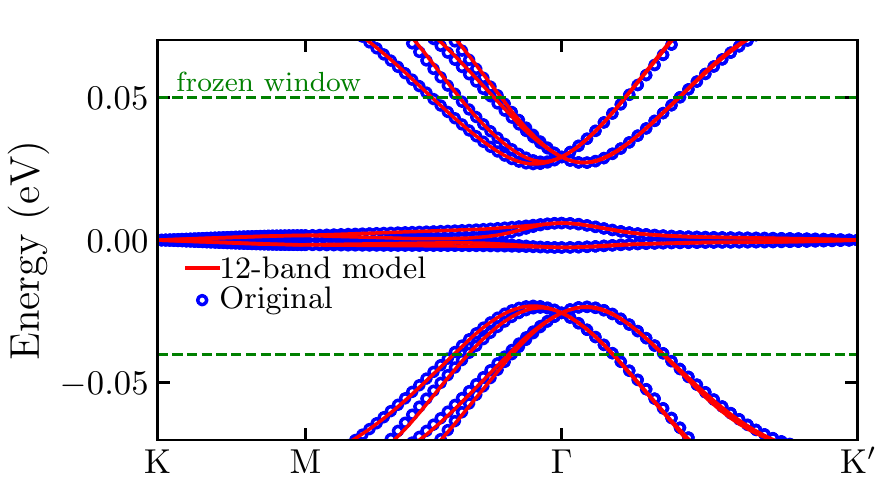}
\caption{Bandstructure computed with twelve-band low-energy Hamiltonian without introducing crystal symmetry constraint, in comparison with the original SK bandstructure.\label{fig:bandstructure_nosym}}
\end{figure}

\begin{figure}
\includegraphics[width=0.48\textwidth]{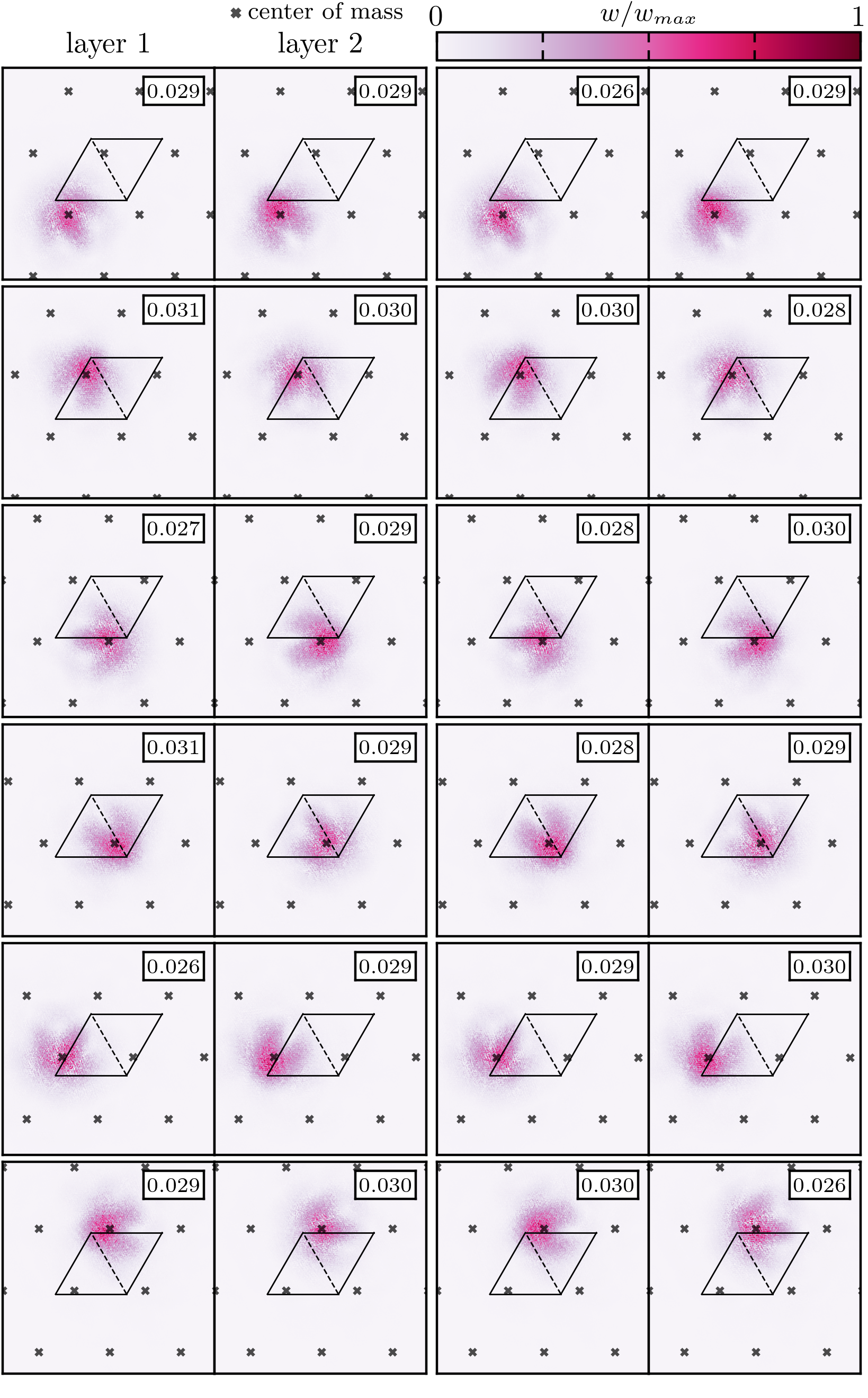}
\caption{The same as Fig.~\ref{fig:wannier_wf_12bands}, but taken from the non-symmetric Wannierization. \label{fig:wannier_wf_12bands_nosym}} 
\end{figure}

We have found out that a more compact twelve-band low-energy TB  model for magic-angle TBG can be obtained by switching off the crystal symmetry constraint and using the frozen window technique in the disentanglement step. The resulting bandstructure together with frozen window used in the calculation is shown in the Fig~\ref{fig:bandstructure_nosym}. In this case the flat bands are reproduced more accurately in comparison with the crystal-symmetry constrained calculation, while having larger error in other bands, which are outside the frozen window. One can also see from the Wannier functions plot of Fig.~\ref{fig:wannier_wf_12bands_nosym} that centers of masses of each orbital form the Kagome lattice, and when plotted all together, form two hexagons per unit cell sharing one edge. Such calculations is, however, very sensitive to both trial orbital set $\omega_n$ and the choice of the frozen window. In the particular calculation of this section the trial orbitals were taken as lobes of trial orbitals for four-band model (i.e., taking separately each of three pockets from functions similar to ones in Fig.~\ref{fig:wfs_4bands})

\section{TR symmetry constraint in the disentanglement algorithm}
In the disentanglement procedure, the  Eq.~\eqref{eq:UD_eq_dU} is solved for $V^\kk$ instead of $U^\kk$ at every iteration of disentanglement algorithm. In order to ensure that these matrices obey time-reversal symmetry, one should incorporate the insertion of Eq.~\eqref{eq:TRsymmetrize_U_isolated} to be applied self-consistently during the solution of Eq.~\eqref{eq:UD_eq_dU}. Instead, we have implemented this insertion to be applied only after solving Eq.~\ref{eq:UD_eq_dU} at every disentanglement iteration step. This gave slightly slower iterative convergence of the disentanglement algorithm with, however, an advantage of simpler implementation. Note that such approach can be extended to a spinor case.

\begin{figure}
\includegraphics[width=0.48\textwidth]{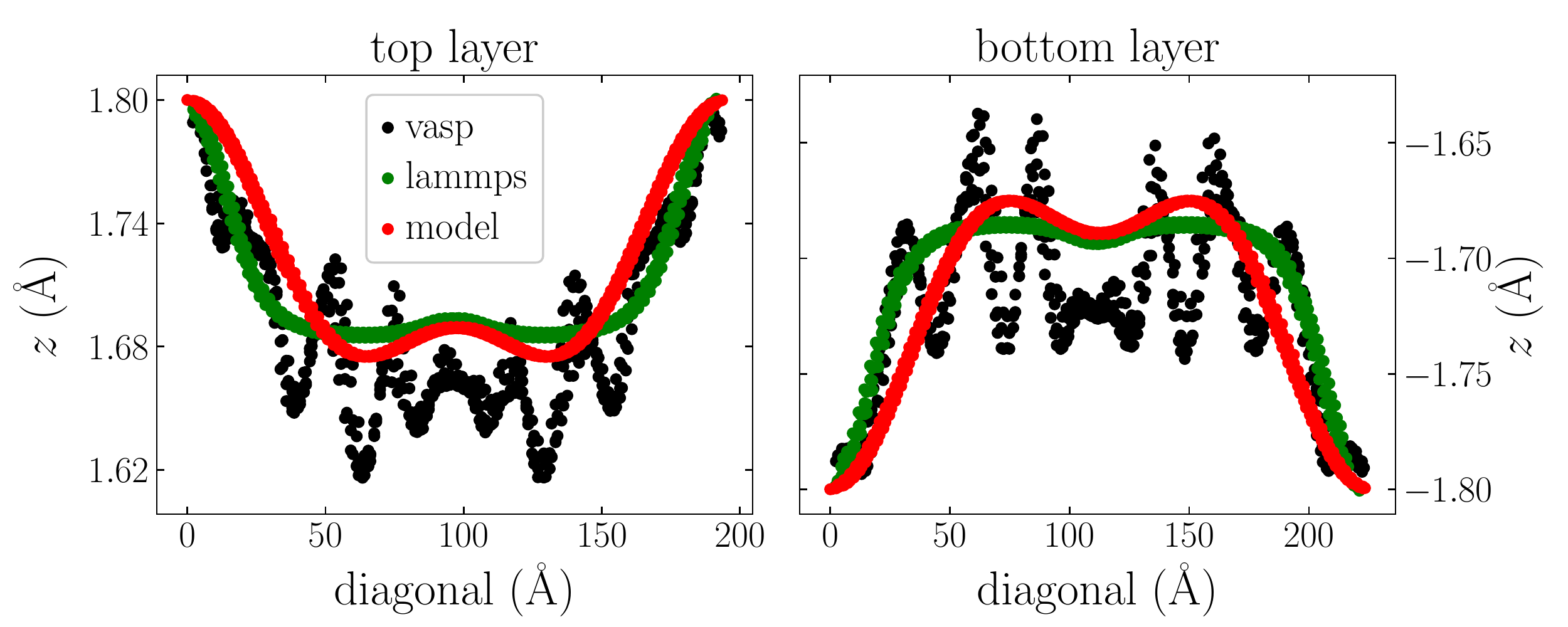}
\caption{Top layer (left) and bottom layer (right) vertical coordinate of atoms in vicinity of the long diagonal of the TBG UC in three different crystal structures. The model structure is given by the Eq.~\eqref{eq:corrugation}, the structure relaxed using classical forces (lammps code) was given by the authors of Ref.~\cite{Haddadi2020}, while the DFT-relaxed structure (VASP code) was taken from Ref.~\cite{Cantele2020}.\label{fig:compareCorrugation}} 
\end{figure}

\begin{figure}
\includegraphics[width=0.48\textwidth]{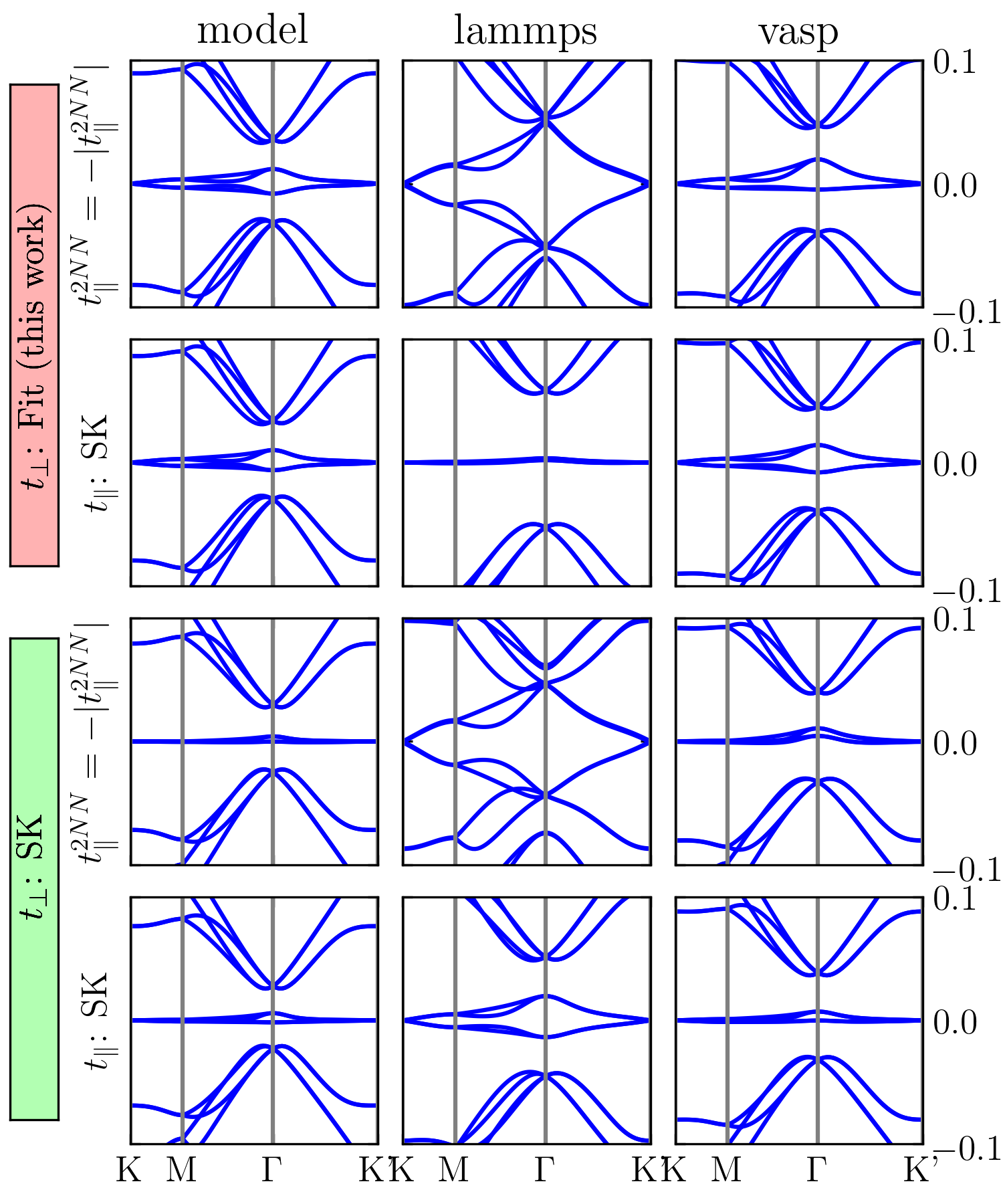}
\caption{Electronic bands computed with various in-plane and out-of-plane TB model parametrizations for $1.08^\circ$-TBG. The first two rows are obtained with the SK parametrization for $t_\perp$ developed in this work (Sec.~\ref{sec:extraction_hopping}), while the standard SK parameters for $t_\perp$ were used in the lower two rows. The standard SK parameters for $t_{\|}$ were used in the second and the fourth row, while the first and the third row correspond to using the ab-initio TB parameters for SLG for in-plane Hamiltonian of TBG with, however, reversed sign of the second-nearest neighbor hopping amplitude. Column names are explained in the caption of Fig.~\ref{fig:compareCorrugation}, and all vertical energy axes are in eV. Finally, values for the standard SK parameters can be found in Fig.~\ref{fig:tb_parametrization} of the main text and in references of its caption.\label{fig:compareCorrugationBands}} 
\end{figure}

\begin{figure}
    \includegraphics[width=0.48\textwidth]{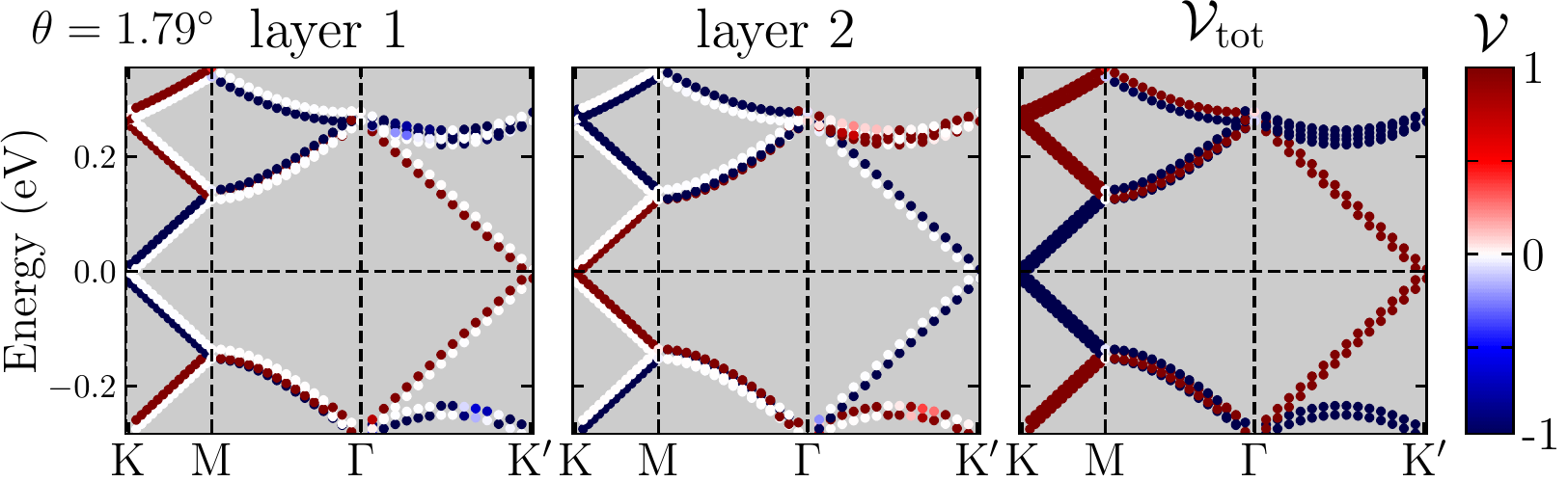}
    \caption{Valley- and layer-resolved bandstructure of $1.79^\circ$ TBG ignoring the out-of-plane coupling in the tight-binding model. The $\mathcal{V}_{tot}$ is a difference between valley characters ($\mathcal{V}$) of two layers. Note, that odd and even bands are shifted vertically with respect to each other for a better observation of the valley character, which is normalized to span $[-1,1]$ range.}
    \label{fig:valle_noTperp}
\end{figure}
\begin{figure}
    \includegraphics[width=0.48\textwidth]{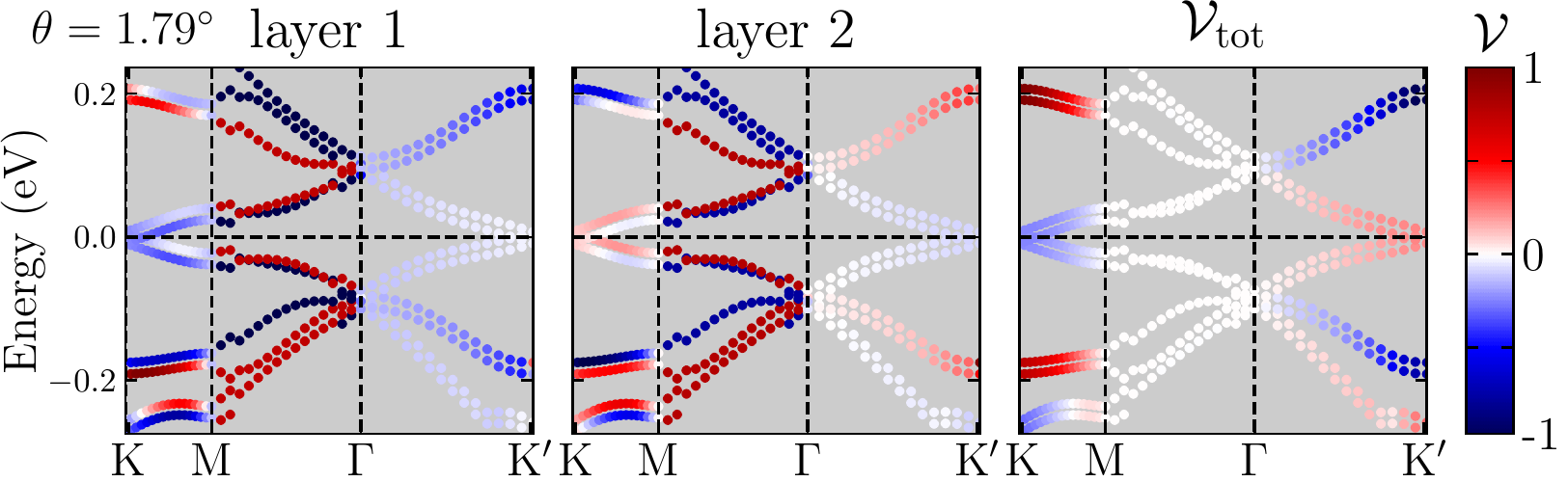}
    \caption{The same as Fig.~\ref{fig:valle_noTperp}, but with finte $t_\perp$ parametrization used in our work.}
    \label{fig:valle_Tperp}
\end{figure}

\section{Corrugations in the $1.08^\circ$-TBG\label{sec:compareCorrugations}}
In this section we compare different relaxed atomic structures for $1.08^\circ$-TBG, and present the electronic structures obtained by different TB Hamiltonians. The model of corrugation used in this work [Eq.~\eqref{eq:corrugation}] is basically the Fourier expansion of the DFT-relaxed structures at twist angles larger than $2^\circ$~\cite{Uchida2014}. It was shown recently~\cite{Cantele2020} that $1.08^\circ$ DFT-relaxed TBG possesses a crystal structure, which has also high-frequency modulations of the carbon's $z$-coordinates with respect to the in-plane location, as can be seen in the Fig.~\ref{fig:compareCorrugation}. We find, however, that the electronic bandstructures computed with various TB Hamiltonians are similar in these two cases (first and last columns of Fig.~\ref{fig:compareCorrugationBands}). On the other hand, the crystal structure obtained with the classical force field methods gives quite different electronic bandstructures except the pure SK case with original parameters (last row of Fig.~\ref{fig:compareCorrugationBands}). Taking the DFT-relaxed crystal structure as the reference one, we assume that the right SK parameters must perform the best in this case. We see from Fig.~\ref{fig:compareCorrugationBands} that parameters used in this work (corresponding to the second row of the figure) have satisfactory performance in flat-bands bandwidths ($\sim20$ meV) and gaps between narrow and other bands ($\sim30$ meV) against the corresponding experimental values of $10$ and $30-60$ meV~\cite{Cao2018_TBG_superconductivity} (for $1.05^\circ$-TBG) correspondingly (keep in mind the reduction by a factor of two of the narrow bands bandwidth when reducing the twist angle to $1.05^\circ$). The parameters from the first row can, in principle, be used as well, but the ad-hoc change of the sign of the in-plane second-nearest hopping amplitude has no justification, therefore, the conventional in-plane SK-TB Hamiltonian was chosen for our calculations.

\section{Valley projection}
We have found out that the out-of-plane coupling in the tight-binding model gives an essential valley mixing in TBG. These can be seen from the valley and layer projected bandstructures, which can be obtained by evaluating the projector operator from Ref.~\cite{Ramires2019,Colomes2017} on TBG eigenfunctions across the Brillouin zone. For example, when the out-of-plane $t_\perp$ in the tight-binding model is off, we see a robust two-valley structure in each of layers in Fig.~\ref{fig:valle_noTperp}. In contrast, the finite $t_\perp$ makes the bands being heavily mixed in the valley character (Fig.~\ref{fig:valle_Tperp}). In addition to that, the Wannierization algorithm do not employ a valley character resulting in the fact that the Fourier transform step of Eq.~\eqref{eq:Wannier_orbital_from_Bloch} picks up bands of one valley character at K and of the opposite one at K$'$, which contributes to zero valley character at a Wannier orbital. The matrices $U^{\kk}$ in Eq.~\eqref{eq:Wannier_orbital_from_Bloch} could, potentially, sort the bands according to valley character, but this would be an additional constraint for Wannierization, which can be, in principle, implemented. Note, that the conserved quantum number is $\mathcal{V_\mathrm{tot}}=\mathcal{V}_1-\mathcal{V}_2$~\cite{Ramires2019}, i.e., difference of valley characters between two layers. If such operation is performed, the resulting $\mathcal{V_\mathrm{tot}}$ behaves smoothly across the Brilloiun zone, and  $\mathcal{V_\mathrm{tot}}$ on $\Gamma$-M path becomes zero in the finite $t_\perp$ case. With vanishing $t_\perp$, the $\mathcal{V_\mathrm{tot}}$ is either $1$ or $-1$ except highly symmetric $\Gamma$ and M points, where it becomes zero.
\end{document}